\documentclass[iop,apj,numberedappendix,twocolappendix,revtex4]{emulateapj}  
\usepackage{amsmath}
\usepackage{bm}
\usepackage{graphicx}
\usepackage[pdfa]{hyperref}
\usepackage{natbib}
\usepackage{color}
\hypersetup{colorlinks,
linkcolor=blue,
citecolor=blue,
filecolor=blue,
urlcolor=blue}

\usepackage{enumerate}
\usepackage{textcomp}

\usepackage{multirow}
\slugcomment{}

\begin{document}
\title{Post-Newtonian Jeans analysis}
\shorttitle{Post-Newtonian Jeans analysis}
\author{Elham Nazari }
\author{Ali Kazemi }
\author{Mahmood Roshan \altaffilmark{*}}
\author{Shahram Abbassi }
\affil{Department of Physics, Ferdowsi University of Mashhad, P.O. Box 1436, Mashhad, Iran; \textcolor{blue}{mroshan@um.ac.ir}}

\begin{abstract}
The Jeans analysis is studied in the first post-Newtonian limit. In other words, the relativistic effects on the local gravitational instability are considered for systems where characteristic velocity of the system and corresponding gravitational field are higher than what permitted in Newtonian limit. The dispersion relation for propagation of small perturbations is found in the post-Newtonian approximation using two different techniques. A new Jeans mass is derived and compared to the standard Jeans mass. In this limit, the relativistic effects make the new Jeans mass to be smaller than the Newtonian Jeans mass. Furthermore, the fractional difference between these two masses increases when temperature/pressure of the system increases. Interestingly, in this limit pressure can help the gravitational instability instead of preventing it.  Finally the results are applied to high temperature astrophysical systems and the possibility of local fragmentations in some relativistic systems is investigated.
\end{abstract}
\section{Introduction}
\label{intro}
One of the key factors for star formation and fragmentation in the interstellar gas clouds is the gravitational instability or the so-called Jeans instability. In fact, Jeans instability in non-rotating environment happens when the internal gas pressure can not prevent the gravitational collapse. In this case, perturbations with masses larger than the Jeans mass will collapse. In other words, small perturbations are amplified when their scales are larger than a specific length, i.e.,  the Jeans length. This instability has been widely investigated in different situations by taking into account various effects. In the rotating mediums, finding a criterion for Jeans instability is much more complicated. In this case the stabilizing effects of the angular momentum of the system should be taken into account. For example, the Jeans analysis on the surface of rotating disk galaxies leads to the so-called Toomre criterion, \cite{toomre1964gravitational}. This criterion has been modified to include the effect of the thickness of the disk, see \cite{toomre1964gravitational} and \cite{vandervoort1970density}. Also Toomre criterion for multicomponent galactic disk has been studied in \cite{kato1972oscillation} , \cite{bertin1988global} , \cite{romeo1992stability} , \cite{wang1994gravitational} and \cite{romeo2011effective} , \cite{shadmehri2012gravitational} , \cite{rafikov2001local} , \cite{jog1984two} , \cite{elmegreen1995effective} and \cite{jog1996local} . \cite{gammie1996linear} has also considered the effect of viscosity to the local gravitational stability. Gravitational instability in the presence of an external tidal field has been studied in \cite{jog2013jeans} .

Furthermore the local stability in two-component galactic disk with gas dissipation has been studied by \cite{elmegreen2011gravitational} . In magnetized regions, the magnetic tensions counters the stabilizing effect of the Coriolis force, and allows the self-gravitating collapse of the overdense regions (\citealp{elmegreen1987supercloud,kim2001amplification}). This instability is known as the magneto-Jeans instability. 

The Jeans analysis has also been investigated for filamentary structures. Theses structures reveal in observations and simulation. Consequently interest in the growth of small perturbations within filamentary systems has raised in recent years. Jeans analysis of these systems helps to study their local and global stability, for a brief review of the relevant literature and recent works on the Jeans analysis of cylindrically symmetric configurations we refer the reader to \cite{freundlich2014local}, \cite{hosseinirad2017gravitational}.

The Jeans instability has been also studied in the context of some gravitational theories in which the gravitational law is different from  the standard case. In these theories the Jeans mass, in principle, is different from the Newtonian case and consequently the growth rate and the dispersion relation of the small perturbations are different from Newtonian gravity. These differences in some astrophysical system may lead to observable differences to discriminate between gravity theories, for example see \cite{roshan2014jeans,roshan2015local,roshan2015stability}, \cite{capozziello2012jeans} .

In this paper, we study the relativistic effects on the Jeans instability. More specifically using the post-Newtonian (PN) theory, we find the Jeans mass and the stability criterion for astrophysical systems where the characteristic velocity of the fluid and the gravitational forces throughout the system are higher than what allowed in the Newtonian regime. This is the case, for example, near the galactic nucleus or accretion disks around black holes (BHs). For example we apply the results to hyper massive neutron stars and Neutrino dominated accretion disks and find some significant deviations from the standard Jeans analysis.

 Relativistic effects on the different kinds of instabilities are of interest since they can significantly influence the dynamics of the background system. For example in \cite{siegel2013magnetorotational} it has been shown that magnetorotational instability has a profound impact on the evolution of the hypermassive neutron stars. The local stability of strongly magnetized relativistic tori orbiting Kerr black holes, has been investigated in \cite{wielgus2015local} for the case of a purely toroidal magnetic field topology. They showed that such tori are subject to an unstable non-axisymmetric magnetorotational mode.

 For some works on the relativistic Kelvin-Helmholtz instability we refer the reader to \cite{blandford1976kelvin,ferrari1978relativistic,hardee1988spatial,zhang2009three} . Furthermore the dynamical instability of inspiraling neutron-star binaries near coalescence in the PN limit has been studied in \cite{lai1996innermost,faber2000post}. 

It is important mentioning that in general relativity field equations can be solved analytically only for some restricted cases where the matter distribution possesses special symmetries. On the other hand self-gravitating systems in reality do not possess such perfect symmetries. As a result, beside numerical simulations, approximative methods are used to solve Einstein's equations. One of the powerful approaches is the PN theory (\citealp{chandrasekhar1965post,chandrasekhar1967post,chandrasekhar1969conservation,chandrasekhar1969Second,chandrasekhar1970212}). This theory works in systems where the gravitational field is suitably weak and motions inside the matter source also are appropriately slow compared with the speed of light.  But both velocity and gravitational field are high enough to lie outside the realm of validity of Newtonian description. For some applications of this theory we refer to: the equation of motion of binary pulsars (\citealp{blandford1976arrival,epstein1977binary,hulse1975Discovery,damour1991orbital}), tests of general relativity in solar system (\citealp{thorne1987300,will1994proceedings}), and gravitational radiation reaction (\citealp{chandrasekhar1970212,burke1971gravitational,blanchet2006gravitational}). Also it has been used to study the relativistic effects in accretion discs around BHs (\citealp{demianski1997dynamics}).

There are two different approaches to PN theory. One of these derivations, which is referred as the classic approach to PN theory, is based on the standard formulation of the Einstein field equations. An alternative derivation, the modern approach, of PN theory is based on the Landau-Lifshitz formulation of Einstein field equations. In this method, the PN metric is restricted to a near zone\footnote{Near zone is a three dimensional region which the characteristic length scale is smaller than $\lambda_\text{c}$, where $\lambda_\text{c}$ is the characteristic wavelength of the gravitational radiation.} within one characteristic wavelength of the radiation, while there is not any clear restriction in the classic approach. 
 For an excellent introduction to PN theory we refer the reader to \cite{poisson2014gravity} where modern and classical formulations of the theory are introduced; and to \cite{will2014confrontation} for a comprehensive review of the applications of this approach. It is important to mention that modern approach possesses some important advantages and removes several ambiguities in the classic approach. Here, we use the modern approach as well as the classic approach to investigate the local gravitational stability.

Naturally approximated methods for solving Einstein's equations have some restrictions . More specifically, as we mentioned before, the PN approximation is limited to weak fields as well as slow motions. However recent words claim that PN approximation is unreasonably successful to describe systems which are not in the PN approximation(\citealp{will2011unreasonable}).

 The layout of the paper is the following. In Sec.\ref{hydrodynamics}, we briefly review the equations of the PN hydrodynamics. Also we derive the dispersion relation for propagation of small perturbations in PN approximation using two different approaches. In Sec. \ref{jeans analysis} we find a new Jeans mass, i.e., the PN Jeans mass and in Sec. \ref{Astro sec} we apply it to some high temperature systems. We also compare it with the standard Jeans mass.
 Finally in Sec. \ref{Conclusion}, results are discussed.

\section{Equations of post-Newtonian hydrodynamics  }\label{hydrodynamics}
Before  moving on to introduce the PN hydrodynamics, it is important to mention that we assume a perfect fluid energy-momentum tensor for our fluid systems throughout this paper. In other words dissipative effects have not been considered in our analysis. 

In this section we briefly review the PN hydrodynamics.  As mentioned in the introduction section, the realm of validity of the PN theory is restricted to systems where the gravitational fields are weak and the motions (all the motions within the matter distribution) are slow compared with the speed of light. Keeping in mind these assumptions, the space-time metric of a PN system can be obtained. For a comprehensive review of the subject we refer the reader to \cite{poisson2014gravity}. The resulting components of the PN metric are given by
\begin{equation}
\begin{split}
&g_{00} = -1+\frac{2}{c^2}U+\frac{2}{c^4}(\Psi-U^2)+O(c^{-6})\\&
g_{0j} = -4/c^3U_j+O(c^{-5})\\&
g_{jk}=\delta_{jk} \left( 1+\frac{2}{c^2}U\right) +O(c^{-4})
\end{split}
\end{equation}
where $\Psi$ is defined as
\begin{equation}\label{big Psi}
\Psi = \psi+\frac{1}{2}\partial_{tt}X
\end{equation}
The potentials that appear in the metric are defined by the following differential equations
\begin{equation}\label{poisson u}
\nabla^2U = -4\pi G\rho^*
\end{equation}
\begin{equation}\label{poisson x}
\nabla^2X = 2U
\end{equation}
\begin{equation}\label{poisson psi}
\nabla^2\psi = -4\pi G\rho^* \left( \frac{3}{2} v^2-U+\Pi+\frac{3p}{\rho^*}\right) 
\end{equation}
\begin{equation}\label{poisson vecu}
\nabla^2\bm{U} = -4\pi G\rho^*\bm{v}
\end{equation}
Where $U$ is the Newtonian potential, $\bm{U}$ is the vector potential, and $X$ is the superpotential which its source term $2 U$ extends over all space.
In fact these equations are replaced with the Poisson equation in the Newtonian gravity. These differential equations can be simply solved, namely
\begin{equation}\label{po in}
\begin{split}
&U(t,\bm{x}) = G\int \frac{\rho^{*'}}{\left| \bm{x-x'}\right| } d^3x'\\&
\psi(t,\bm{x}) = G \int \frac{\rho^{*'} (\frac{3}{2}v^{'2}-U'+\Pi'+3p'/\rho^{*'})}{\left|\bm{ x-x'}\right| }\\&
X(t,\bm{x}) = G \int \rho^{*'} \left| \bm{x-x'}\right|  d^3x'\\&
U^j(t,\bm{x}) = G \int \frac{\rho^{*'}v^{'j}}{\left|\bm{ x-x'}\right| } d^3x'
\end{split}
\end{equation}
in which the primed fluid variables are evaluated at time $t$ and position $\bm{x'}$.
It should be noted that the modern approach clarifies these potentials in terms of near-zone integrals. It means that the field point $x$ is within the near zone, but in the classic approach there is no such restriction on domain of the field point. In addition, in modern approach the domain of these integrals, i.e., $\mathcal{M}$, is a surface of constant time which is bounded by a sphere of radius $\mathcal{R}\sim \lambda_c$, while in the context of classic approach, it depends on the boundary of the potential sources. Therefore, some potentials with non-compact sources (e.g. the superpotential $X$) are ill-defined and the integral representations of them are divergent. Consequently, the field equations \eqref{poisson u}-\eqref{poisson vecu} may lead to ambiguous solutions in this approach, for more details see \citet{poisson2014gravity}.

The matter variables in the PN approach are $  \left\lbrace  \rho^*,p,\Pi,\bm{v}  \right\rbrace $ where $\rho^*$ is the conserved mass density defined as
\begin{equation}
\rho^* = \sqrt{-g}\gamma \rho = \sqrt{-g} \rho u^0/c
\end{equation}
in which $\rho$ is the proper density and $g$ is the determinant of the metric tensor. Also $u^0$ is the zeroth component of the velocity four-vector. It is clear that in the Newtonian limit there is no distinction between $\rho^*$ and the proper density $\rho$.
furthermore $p$ is the pressure, $\Pi=\epsilon/\rho^*$ is the internal energy per unit mass, $\epsilon$ is the proper internal energy density and $\bm{v}$ is the fluid's velocity field, and is defined with respect to the time coordinate $t$.  

Now let us introduce the hydrodynamics equations in the first PN (1\tiny PN\normalsize) approximation. In fact, the Newtonian equations of hydrodynamics change because of both special relativistic and general relativistic effects. In this case the continuity equation can be written as
\begin{equation}\label{cont}
\partial_t\rho^* + \partial_j(\rho^*v^j) = 0
\end{equation}
where $\partial_{\alpha}=\frac{\partial}{\partial x^{\alpha}}$.
It is obvious that PN continuity equation is similar to the standard case and the only difference is the appearance of $\rho^*$ instead of $\rho$. Where the energy density $\rho$ and $\rho^*$ are related as 
\begin{equation}\label{rho star}
\rho = \rho^* \left( 1-\frac{v^2}{2c^2}-\frac{3U}{c^2}\right) +O(c^{-4})
\end{equation}
 Also we assume that the fluid is in local thermodynamic equilibrium. In this case the exact statement of the first law of thermodynamics for the perfect fluids is $d \Pi= (p/\rho^2) d\rho$. Consequently we can write
\begin{equation}\label{first law}
\frac{d\Pi}{dt} = \frac{p}{\rho^{*2}}\frac{d\rho^*}{dt}+O(c^{-2})
\end{equation}
where $d/dt = \partial_t+v^k\partial_k$ is the Lagrangian time derivative. On the other hand, the PN Euler's equation can be written as
\begin{eqnarray}\label{euler}
\begin{split}
 & \rho^*\frac{dv^j}{dt} =  -\partial_jp+\rho^*\partial_jU 
 +\frac{1}{c^2}\Big[ \left( \frac{1}{2}v^2+U+\Pi+\frac{p}{\rho^*}\right) 
 \partial_jp\\&  -v^j\partial_tp\Big]
 +\frac{\rho^*}{c^2}\Big[ (v^2-4U)\partial_jU-v^j(3\partial_tU+4v^k\partial_kU)
\\&+ 4\partial_tU_j+4v^k(\partial_kU_j-\partial_jU_k)+\partial_j\Psi\Big]+O(c^{-4})
 \end{split}
\end{eqnarray}
Furthermore the energy equation in the PN limit takes the following form
\begin{equation}\label{ene}
\rho^*\partial_t (\frac{1}{2}v^2+\Pi) + \rho^*\bm{v}.\nabla (\frac{1}{2}v^2+\Pi)+\nabla.(p\bm{v})-\rho^*\bm{v}.\nabla U=0
\end{equation}
After some manipulations one can verify that, in the case of perfect fluids, the above equation is the same as the first law of thermodynamics. So one can recognize it as a statement of conservation of energy for isentropic flows. Therefore, equations (\ref{poisson u})-(\ref{poisson vecu}), (\ref{cont})-(\ref{euler}) 
and an equation of state make a complete set of differential equations for describing a PN perfect fluid. 
\subsection{Gravitational instability criterion in the post-Newtonian limit}
A common way to derive the Jeans mass in Newtonian gravity is to use the virial theorem. In the PN limit one may use the PN virial theorem, see (\citealp{roshan2012parametrized,chandrasekhar1965post}) for virial theorem in the PN approximation, in order to find the PN version of the Jeans mass. However, here we follow another approach by studying the linear perturbation analysis of the self-gravitating system.

In order to have a crude estimate for PN overall effects on the Jeans stability, let us study the gravitational binding energy of a spherically symmetric object in this approximation. The general form of the binding energy of a spherically symmetric distribution of matter in general relativity is given by \cite{hobson2006general} 
\begin{equation}
E_{\text{B}}=(\tilde{M}-M)c^2
\end{equation}
where $\tilde{M}$ and $M$ are different masses defined as
\begin{equation}
\begin{split}
&\tilde{M}=4 \pi \int _0^R \rho^* (r)\sqrt{g_{rr}(r)}r^2 dr\\&
M=4 \pi \int _0^R \rho^* (r)r^2 dr
\end{split}
\end{equation}
where $R$ is the radius of the system and $g_{rr}$ is the $rr$ component of the metric. In particular, one can easily show that the gravitational binding energy of a spherical symmetric and static fluid, which has constant density and extends to coordinate radius $r=R$, can be written in the form
\begin{equation}
E_{\text{B}}=\frac{16 \,\pi^2}{15}G \rho^{*2}R^5
\end{equation}
Now to find the PN corrections to the binding energy we substitute equation (\ref{rho star}) into above equation. Finally we find the PN gravitational binding energy $E_{\text{Bp}}$
\begin{equation}
E_{\text{Bp}}\simeq E_{\text{BN}}+\frac{32 \,\pi^2}{5\,c^2}G \rho^{2}U\,R^5+O(c^{-4})
\end{equation}
where $E_{\text{BN}}$ is the binding energy of a uniform sphere in Newtonian gravity and is equal to $\frac{16 \,\pi^2}{15}G \rho^{2}R^5$. It is clear that $E_{\text{Bp}}>E_{\text{BN}}$. This means that PN effects increase the gravitational binding energy of this system. Therefore for a system deep in the PN limit, keeping in mind the virial theorem, one may conclude that more random motion is required to hold the system in an equilibrium state. In other words it seems that PN corrections would support the gravitational instability. We will show that this simple conclusion is true. 

Now, to study the linear perturbation growth in PN limit, we need to find the linearized PN hydrodynamics. Using these equations one can, in principle, find a dispersion relation governing the propagating of the first order perturbations. On the other hand, this dispersion relation helps to find a criterion for stability of the system against small overdensities. This criterion in the Newtonian case is the so-called Jeans stability criterion. Our purpose here is to find the relativistic corrections to it. As we will see, analysis of the stability is more complicated than the Newtonian case. In order to check the reliability of the results, we find the dispersion relation using two slightly different procedures and find a same dispersion relation. 

In the first method, one can solve the linearized equations by a perturbative procedure. To do so, we perturb all the quantities up to the first order of perturbations. Then using the Fourier expansion one can solve the linearized equations and find the Fourier coefficients. Finally the dispersion relation is derived. Let us first assume that the fluid is a barotropic fluid and obeys the following equation of state
\begin{equation}\label{eq state}
p = p (\rho^*)
\end{equation}

Furthermore we assume that every quantity $Q_0(\bm{x})$ is perturbed as $Q(\bm{x},t) = Q_0(\bm{x}) + Q_1(\bm{x},t)$ where zero index refers to the non-perturbed quantities and ``$1$" subscript stands for the corresponding small perturbation, i.e., $\frac{Q_1}{Q_0} \ll 1$. Now we substitute these 
perturbed quantities in the PN hydrodynamic equations
 (\ref{poisson u})-(\ref{poisson vecu}), (\ref{cont})-(\ref{euler}), and (\ref{eq state})
 and keep the first order terms. For simplicity it is assumed here that the background fluid is static, infinite and homogeneous. Therefore $\bm{v}$=0 and as a result $\bm{U}_0$ vanishes. Also homogeneity implies that $\rho^*_0$, $p_0$, $\Pi_0$ and $U_0$ are constant. In this case the linearized form of perturbed continuity equation is
\begin{equation}\label{continuity p}
\partial_t\rho_1^* + \rho_0^* \nabla.\bm{v}_1+\bm{v}_1.\nabla\rho^*_0 = 0
\end{equation}
It is important mentioning that the above mentioned assumptions for the background matter distribution do not satisfy the equations. More specifically, the Euler's equation (\ref{euler}) implies that $\nabla\Psi_0=0$. On the other hand, Possion's equations (\ref{poisson u}) and (\ref{poisson psi}) require that
\begin{equation}
\nabla^2U_0 = -4\pi G\rho_0^*
\end{equation}
  and
 \begin{equation}
 \nabla^2\psi_0 = -4\pi G\rho_0^* \left( -U_0+\Pi_0+\frac{3p_0}{\rho_0^*}\right)
 \end{equation}
One can see only if $\rho_0^*=0$, these results are consistent with each other. This inconsistency can be removed by an $\textit{ad hoc}$ assumption that unperturbed potentials are zero and the Poisson's equations are only able to describe the perturbed fluid. This assumption can be considered as a generalized version of the so-called Jeans swindle in the Newtonian case, see \cite{binney2008galactic}. Therefore, following the Newtonian case, we use the Jeans swindle, and complete the stability analysis.

It is easy to verify that the linearized PN Euler's equation is
\begin{eqnarray}\label{Euler p}
\nonumber\rho_0^*\frac{d\bm{v}_1}{dt} &=& -\nabla p_1+\rho_0^*\nabla U_1+\frac{1}{c^2}\left[ \left(\Pi_0+\frac{p_0}{\rho_0^*}\right)\nabla p_1\right] \\
&&+\frac{\rho_0^*}{c^2}\left[4\partial_t\bm{U}_1+\nabla\Psi_1\right]
\end{eqnarray}
By differentiating equation (\ref{continuity p}) with respect to time and combining the result with the divergence of equation (\ref{Euler p}), we find
\begin{eqnarray}\label{d2rho1}
\frac{\partial^2\rho^*_1}{\partial t^2}&-&\nabla^2p_1+\rho^*_0\nabla^2U_1+\frac{1}{c^2} \left[ \left(\Pi_0+\frac{p_0}{\rho^*_0}\right)\nabla^2p_1 \right] \\\nonumber
&&+\frac{\rho^*_0}{c^2} \left[4\nabla.\partial_t\bm{U}_1+\nabla^2\Psi_1\right]  = 0
\end{eqnarray}
Conveniently we use the Fourier expansion as $Q_1 = Q_a e ^{ i ( \bm{k}\cdot\bm{x}-\omega t)} $ for all perturbed quantities. Finally we find
\begin{eqnarray}\label{coefficients}
& &\omega^2\rho_a-k^2p_a + k^2\rho^*_0 U_a+\frac{1}{c^2}\left[ \left(\Pi_0+\frac{p_0}{\rho^*_0}\right) k^2 p_a\right]\\\nonumber
&&+\frac{\rho^*_0}{c^2} \left[- 4 \omega \bm{k} . \bm{U}_a + k^2\psi_a  + \omega^2 U_a\right] = 0
\end{eqnarray}
note that
\begin{equation}
\nabla^2\Psi_1 =\nabla^2\psi_1+\frac{1}{2}\partial_{tt}\nabla^2X_1 = \nabla^2\psi_1+\partial_{tt}U_1 . 
\end{equation}
Now using the equation of state, Poisson equations and the first law of thermodynamics one can easily find all the Fourier coefficients, i.e., $Q_a$, in terms of $\rho_a$. From equation (\ref{eq state}) we have
\begin{equation}\label{p_a}
p_a = c_s^2\rho_a
\end{equation}
where $c_{\text{s}}$ is the sound speed, i.e., $ c_{\text{s}}^2=\left(\frac{d p }{d \rho}\right)_{\rho_0}$. On the other hand equation (\ref{poisson u}) leads to
\begin{equation}\label{u_a}
U_a=\frac{4\pi G}{k^2}\rho_a
\end{equation}
and equation (\ref{poisson vecu}) gives
\begin{equation}
\bm{U}_a=\frac{4\pi G\rho^*_0}{k^2} \bm{v}_a
\end{equation}
By substituting the perturbed quantities in the mentioned form in to equation  (\ref{Euler p}), one can easily show that $\bm{v}$ and $\bm{k}$ are parallel. As a result by considering equation (\ref{continuity p}) we find
\begin{equation}\label{va}
v_a = \frac{\omega}{k\rho^*_0}\rho_a
\end{equation}
On the other hand using equation (\ref{first law})  we can simply relate $\Pi_a$ to $\rho_a$ as
\begin{equation}\label{Pi_a}
\Pi_a = \frac{p_0}{\rho^{*2}_0}\rho_a
\end{equation}
To find $\psi_a$ with respect to $\rho_a$, let us rewrite equation (\ref{poisson psi}) as
\begin{eqnarray}
k^2\psi_a &=& 4\pi G \left[ \rho^*_0\left( -U_a+\Pi_a+\frac{3p_a}{\rho^*_0}- \frac{3p_0}{\rho^{*2}_0}\rho_a\right)\right.\\\nonumber
&& \left.+\rho_a \left(\Pi_0+\frac{3p_0}{\rho^*_0}\right) \right] 
\end{eqnarray}
Now using equations (\ref{p_a}), (\ref{u_a}) and (\ref{Pi_a}) we have
\begin{equation}\label{psi_a}
\begin{split}
\psi_a = & \frac{4\pi G\rho_a}{k^2}\Big[ \rho^*_0 \left( -\frac{4\pi G}{k^2}-2\frac{p_0}{\rho^{*2}_0}+3\frac{c_s^2}{\rho^*_0}\right)\\&+\left(\Pi_0+\frac{3p_0}{\rho^*_0}\right) \Big]
\end{split}
\end{equation}
Finally by substituting (\ref{p_a})-(\ref{va}) and (\ref{psi_a}) into (\ref{coefficients}) and keeping terms up to $O(c^{-2})$, we find the following dispersion relation
\begin{eqnarray}\label{mr1}
\nonumber\omega^2 &=& c_s^2k^2 - 4\pi G\rho^*_0-\frac{1}{c^2}\left(\Pi_0+\frac{p_0}{\rho^*_0} \right) \left[c_s^2k^2 + 4\pi G\rho^*_0 \right]\\
&& - \frac{32\pi^2 G^2\rho^{*2}_0}{c^2k^2}
\end{eqnarray}
 By ignoring the PN correction terms, i.e., terms including $c^{-2}$ coefficients, the Newtonian dispersion relation is recovered. Equation (\ref{mr1}) is the main result of this section. In fact, as we mentioned before, we will find the PN Jeans stability criterion using this dispersion relation. It is important here to note that dispersion relation \eqref{mr1} is not valid for long wave lengths. One may simply find and constraint on the wave number of the perturbation mode by taking into account that the PN corrections in \eqref{mr1} must be smaller than the Newtonian terms. Therefore we expect that

\begin{equation}\label{condi}
k^2>\sqrt{2}\frac{c_s\rho_0^{*}}{c\rho_0}k_{\text{J}}^2
\end{equation}
where the standard Jeans wavenumber $k_{\text{J}}$ is defined as
\begin{equation}\label{kJN}
k_{\text{J}}^2=\frac{4\pi G \rho_0}{c_s^2}
\end{equation}

Before moving on to find the stability criterion, let us derive the dispersion relation once again using an another procedure based on the modern approach of the PN theory. In this case, it is straightforward to derive the following relations by using the equation of state (\ref{eq state}) and Poisson's equations (\ref{poisson u})-(\ref{poisson vecu})
\begin{equation}
\nabla^2p_1 = c_s^2\nabla^2\rho^*_1
\end{equation}
\begin{equation}
\nabla^2U_1 = -4\pi G\rho^*_1
\end{equation}
\begin{equation}
\nabla^2X_1 = 2U_1
\end{equation}
\begin{eqnarray}
\nonumber\nabla^2\psi_1& =& -4\pi G \left[ \rho^*_1 \left(\Pi_0+\frac{3p_0}{\rho^*_0}\right)\right.\\ &&\left.+\rho^*_0\left( -U_1+\Pi_1+\frac{3p_1}{\rho^*_0}-3\rho^*_1\frac{p_0}{\rho^{*2}_0}\right) \right] 
\end{eqnarray}
\begin{equation}
\nabla^2\bm{U}_1 = -4\pi G\rho^*_0\bm{v}_1
\end{equation}
using these differential equations together with equation (\ref{d2rho1}) one can write
\begin{eqnarray}\label{d2rho tt and n}
\begin{split}
&\frac{\partial^2\rho^*_1}{\partial t^2}-c_s^2\nabla^2\rho^*_1 -4\pi G\rho^*_0\rho^*_1+\frac{\rho^*_0}{c^2}\Big[ \left(\Pi_0+\frac{p_0}{\rho^*_0}\right) \frac{c_s^2}{\rho^*_0}\nabla^2\rho^*_1  \\& -4\pi G\left( \frac{p_0}{\rho^*_0}+\Pi_0+3c_s^2\right) \rho^*_1 \\& +4 \partial_t\nabla \cdot \bm{U}_1 +4\pi G\rho^*_0U_1+\partial_{tt}U_1\Big] =0
\end{split}
\end{eqnarray}
In order to simplify this expression let us obtain $U_1$, $\partial_{tt}U_1$ and $\partial_t\nabla \cdot \bm{U}_1$ in terms of $\rho^*_1$. We begin with $\partial_{tt}U_1$. In this case the first time derivative of $U_1$ is
\begin{eqnarray}
\partial_tU_1(\bm{x},t)& = & G\partial_t\int\rho^*_1(\bm{x}',t) f d^3x'\\\nonumber
&&=G\int\left( \rho^*_1\frac{\partial f}{\partial t}+f\frac{\partial\rho'^*_1}{\partial t}\right) d^3x'
\end{eqnarray}
where $f = \frac{1}{|\bm{x}-\bm{x}'|}$.  Using equation (\ref{continuity p}) and divergence theorem, and keeping in mind that the surface terms are vanished, after some manipulations we obtain
\begin{equation}
\partial_tU_1 = G\int\rho'^*_0 v'^k_1 \frac{(x-x')^k}{|\bm{x}-\bm{x}'|^3}d^3x'
\end{equation}
similarly the second time derivative can be written as
\begin{eqnarray}
\partial_{tt}U_1& =& G\int \left( \rho'^*_0\frac{dv'_{1k}}{dt}  \frac{(x-x')^k}{|\bm{x}-\bm{x}'|^3}\right.\\
\nonumber&&\left.+\rho'^*_0 v'^k_1v'_j\partial'_j \frac{(x-x')^k}{|\bm{x}-\bm{x}'|^3}\right) d^3x'
\end{eqnarray}
It is important mentioning that only first order terms, i.e $O(1)$, can appear in terms including $U_1$. Therefore in the first term of the above equation we use the Euler's linearized equation and keep the first order terms. Also it is clear that the second term is negligible because gives raise to $O(4)$ terms. Finally we can write
\begin{eqnarray}\label{pttu1}
\begin{split}
\partial_{tt}U_1 = G\int &  \rho'^*_0 \Big[ -\frac{c_s^2}{\rho'^*_0}\nabla'\rho'^*_1
+ \\& \nabla'\left( G\int\frac{\rho^*_1(\bm{y},t)}{|\bm{x}'-\bm{y}|}d^3y\right) \Big]\cdot\frac{(\bm{x}-\bm{x}')}{|\bm{x}-\bm{x}'|^3}d^3x'
\end{split}
\end{eqnarray}
 Now let us calculate $\partial_t\nabla \cdot \bm{U}_1$. Using the fact that $\bm{U}_1 = G\int \frac{\rho^*_0 \bm{v}'_1}{|\bm{x}-\bm{x}'|}d^3x'$ one may write
\begin{eqnarray}
\begin{split}
\partial_tU_{1k} & =  G\partial_t\int \frac{ \rho'^*_0 v'^k_1}{|\mathbf{x}-\bm{x}'|}d^3x'\\&
 = G\int \frac{dv'^k_1}{dt}\frac{\rho'^*_0}{|\bm{x}-\bm{x}'|}d^3x'\\&
~~~+ G\int v'^k_1v'_{1j}\partial_{k'}\frac{\rho'^*_0 }{|\bm{x}-\bm{x}'|}d^3x'
\end{split}
\end{eqnarray}
Because perturbation is linear, again the second term vanishes. In this case, by taking a space derivative we find
\begin{eqnarray}
\partial_k\partial_tU_{1k}& =& G\int\rho'^*_0\left( -\frac{c_s^2\partial'_k\rho'^*_1}{\rho'^*_0}\right.\\\nonumber
&&\left.+ G \,\partial'_k\int\frac{\rho^*_1(\bm{y},t)}{|\bm{x}'-\bm{y}|}d^3y\right) \partial_k\frac{1}{|\bm{x}-\bm{x}'|}d^3x'
\end{eqnarray}
or equivalently we can write
\begin{eqnarray}\label{n.ptu_1}
\nabla \cdot \partial_t\bm{U}_1& =& -G\int\rho'^*_0\left[ -\frac{c_s^2\nabla'\rho'^*_1}{\rho'^*_0} \right.\\\nonumber
&&\left.+ \nabla'\int\frac{G\rho^*_1(\bm{y},t)}{|\bm{x}'-\bm{y}|}d^3y\right]\cdot \frac{(\bm{x}-\bm{x}')}{|\bm{x}-\bm{x}'|^3}d^3x'
\end{eqnarray}
Therefore it is clear form equations (\ref{pttu1}) and (\ref{n.ptu_1}) that
\begin{equation}\label{tt and n}
 \partial_{tt}U_1 = -\partial_t\nabla\cdot\bm{U}_1
\end{equation}
Now using Fourier's expansion for $\rho^*_1$ in the form $\rho^*_1(\bm{x},t) = \rho_ae ^{ i ( \bm{k}\cdot\bm{x}-\omega t)} $ and using equation (\ref{tt and n}), equation (\ref{d2rho tt and n}) takes the following form
\begin{eqnarray}\label{3 ints in w eq}
\begin{split}
& \omega^2-c_s^2k^2+4\pi G\rho^*_0 +\frac{\rho^*_0}{c^2}\left[4\pi G \left(\frac{p_0}{\rho^*_0}+\Pi_0 + 3c_s^2\right)\right.\\
&\left.+\left(\Pi_0+\frac{p_0}{\rho^*_0}\right)\frac{c_s^2}{\rho^*_0}k^2-3 i G  c_s^2 I_1+3G^2\rho^*_0 I_2-4\pi G^2\rho^*_0 I_3 \right]= 0
\end{split}
\end{eqnarray}
where integrals $I_i$ are defined as
\begin{equation}\label{ints}
\begin{split}
& I_1=\int_{\mathcal{M}}\frac{e^{-i\bm{k}\cdot(\bm{x}-\bm{x}')}}{|\bm{x}-\bm{x}'|^3}\bm{k}\cdot(\bm{x}-\bm{x}')d^3x'\\&
I_2=\int_{\mathcal{M}}\nabla'\left(\int_{\mathcal{N}}\frac{e^{-i\bm{k}\cdot(\bm{x}-\bm{y})}}{|\bm{x}'-\bm{y}|}d^3y' \right) . \frac{(\bm{x}-\bm{x}')}{|\bm{x}-\bm{x}'|^3}d^3x'\\&
I_3=\int_{\mathcal{M}}\frac{e^{-i\bm{k}\cdot(\bm{x}-\bm{x}')}}{|\bm{x}-\bm{x}'|}d^3x'
\end{split}
\end{equation}
As previously mentioned, in this second derivation we use the modern approach of PN theory. Thus we deal with the near zone and integrals should be taken in this region. The index $\mathcal{M}$ in this relation means that our calculations are restricted to inside the near zone. Here, we can also see one of the ambiguities that appear in the classic approach. In the classic approach the domain of the integrals over non-compact sources extends to infinity and consequently one may easily show that integrals $I_2$ and $I_3$  in \eqref{ints} will diverge. On the other hand in the modern approach, after integrating on a finite region of radius $\mathcal{R}$, we remove the terms that contain $\mathcal{R}$. In fact if one consider the contributions of wave zone to the integrals, then there will be terms containing $\mathcal{R}$ that cancel the terms arising from the near zone contribution and we will not confront infinities.
This cancellation in modern approach provides a type of regularization of ill-defined integrals that improves the problem of divergent and ambiguous solutions of Poisson equations. For more details on this point we refer the reader to \cite{pati2000post}. 

To solve the integrals, in the spherical coordinate system, without loss of generality, we impose that $\bm{k}$ is in the $z$ direction, i.e  $\bm{k}=k \hat{z}$ where $\hat{z}$ is the unit vector in the direction of $z$ axis. Then integrating over all directions $\bm{x}$, we obtain the following result for $I_2$ 
\begin{equation}
\begin{split}
I_2=\left(\frac{4\pi}{k^2}-\frac{4\pi\cos(k\mathcal{R})}{k^2} \right)\left(4\pi+i\frac{\sin(k\mathcal{R})}{k\mathcal{R}}\right) 
\end{split}
\end{equation}
and similarly for $I_3$ we obtain
\begin{equation}
\begin{split}
I_3 =\frac{4\pi}{k^2}-\frac{4\pi\cos(k\mathcal{R})}{k^2}
\end{split}
\end{equation}
It should be noted that $I_1$ does not diverge and can be easily integrated, the result is $-4 \pi i$. Based on our discussions on the modern approach to PN theory, we remove all terms that depend on $\mathcal{R}$. Therefore one may conclude that $I_2=16\pi^2/k^2$ and $I_3=4\pi/k^2$. Substituting these results into equation \eqref{3 ints in w eq} we find the same dispersion relation obtained in \eqref{mr1}. 

\section{Post-Newtonian Jeans mass}\label{jeans analysis}
So far we have derived the dispersion relation for propagation of the local overdensities in an infinite homogeneous fluid system. If we disregard the 1\tiny PN\normalsize terms, then the equation (\ref{mr1}) coincides with the corresponding dispersion relation in the Newtonian theory. Let us define the following parameters in order to simplify the dispersion relation
\begin{equation}
\begin{split}
& \alpha =\frac{1}{c^2} (\Pi_0+\frac{p_0}{\rho_0^*})<1\\&
c_{\text{sp}}^2 =(1-\alpha )c_{\text{s}}^2\\&
G_\text{p} = (1+\alpha )G
\end{split}
\end{equation}
where $\alpha>0$, $c_{\text{sp}}$ is an effective sound speed and $G_p$ is an effective gravitational constant. As expected in the Newtonian regime $\alpha\rightarrow 0$ and these parameters coincide with $c_{\text{s}}$ and $G$ respectively. The $\text{p}$ index represents the constants in PN theory and we use this notation in the rest of this paper. Generally speaking, the PN effects, effectively, decrease the sound velocity and increase the gravitational constant. Therefore, noting that sound speed and gravitational constant are the representatives of the pressure support and gravitation strength respectively, one may naturally expect that PN effects make the system more unstable against the local perturbations. However, let us find the exact form of the stability criterion in this limit. Using the new variables we rewrite equation (\ref{mr1}) in the following simple form
\begin{equation}\label{w khola3}
\omega^2 = c_{\text{sp}}^2k^2-4\pi G_\text{p}\rho_0-\frac{32\pi^2G^2\rho^{2}_0}{k^2c^2}
\end{equation}
It should be mentioned that since $v=0$ and $U=0$ (by Jeans swindle) one may infer $\rho_0^{*}=\rho_0$. Since $c_{\text{sp}}<c_{\text{s}}$, $G_{\text{p}}>G$ and the third term in \eqref{w khola3} appear with negative sign, the instability width will extend and include a larger interval of wavenumbers. On the other hand a perturbation mode $k$ that is stable in both Newtonian and PN regimes, will have a smaller oscillation frequency in the PN limit. Equivalently for an unstable mode, as expected, the growth rate is larger in the PN case.

By setting $\omega=0$ in equation (\ref{w khola3}), we can obtain the border of stability. In this case the Jeans wavenumber in the PN limit can be obtained as
\begin{equation}\label{kj}
k_{\text{Jp}}^2\simeq k_{\text{J}}^2\left(1+\frac{2}{c^2}\left(c_{\text{s}}^2+\frac{p_0}{\rho_0}+\Pi_0\right)\right)
\end{equation}
Where we have expanded the result and kept only the 1\tiny PN\normalsize corrections. System is stable against perturbation with wavenumber $k$ if $k>k_{\text{Jp}}$, or equivalently if the corresponding wavelength $\lambda$ satisfies $\lambda < \lambda_{\text{Jp}}$, where $\lambda_{\text{Jp}}=2\pi/k_{\text{Jp}}$. It is obvious that $k_{\text{Jp}}>k_{\text{J}}$.  
This explicitly means that perturbations with wavelength in the interval $\lambda_{\text{Jp}}< \lambda <\lambda_{\text{J}}$, which are stable in the Newtonian dynamics, are unstable in the PN regime. In the following we will show that these modes can be excited at mediums with high temperature and low mean molecular weight.

On the other hand, we have another constraint on $k$ in equation (\ref{condi}) to ensure the validity of the PN approximation. Therefore combining these constraints, one may conclude that the perturbation with wavenumber $k$ is unstable provided that
 \begin{equation}\label{mJJP}
\left(\frac{\sqrt{2}c_{\text{s}}}{c}\right)^{\frac{1}{2}}k_{\text{J}}<k<k_{\text{Jp}}
 \end{equation}

It is also instructive to calculate the Jeans mass in the PN limit. In fact, because of the mass-energy equivalence principle, we introduce two types of Jeans masses in the PN limit, one of them constructed from matter density $\rho_0$ and one constructed form matter-energy density $\epsilon_0=(1+\frac{\Pi_0}{c^2})\rho_0$. We call the later case ``Jeans mass-energy". Our purpose is to compare the Jeans mass and the Jeans mass-energy in PN limit with the standard Jeans mass. To do so, we define the Jeans mass $m_{\text{Jp}}$ as the mass inside a sphere with diameter $\lambda_{\text{Jp}}$.
\begin{equation}\label{mJp}
m_{\text{Jp}}\simeq m_{\text{J}}\left(1-\frac{3}{c^2}\left(c_{\text{s}}^2+\frac{p_0}{\rho_0}+\Pi_0\right)\right)
\end{equation}
where the standard Jeans mass $m_{\text{J}}$ is given by
 \begin{equation}\label{mJN}
 m_{\text{J}} = \frac{\pi\rho_0\lambda^3_{\text{J}}}{6}
 \end{equation}
As expected the PN Jeans mass is smaller than the standard case. Therefore, if the mass of a small perturbation is larger than the new Jeans mass $m_{\text{Jp}}$, the system is gravitationally unstable and collapses under its own gravity. Although reduction in the Jeans mass with internal energy is not strange, reduction in Jeans mass with increasing the sound speed and $\frac{p_0}{\rho_0}$ is surprising. In fact in the Newtonian case these quantities work against collapse. Consequently, one may conclude that in the PN regime pressure produces gravity. As we know it is a well-known fact in the context of general relativity that pressure can act as gravity, and here we see this fact even in the 1\tiny PN\normalsize approximation. 
The Jeans mass-energy $M_{\text{Jp}}$ is defined as the mass-energy inside a sphere with diameter $\lambda_{\text{Jp}}$
\begin{equation}\label{eJp}
M_{\text{Jp}}\simeq m_{\text{J}}\left(1-\frac{3}{c^2}\left(c_{\text{s}}^2+\frac{p_0}{\rho_0}+\frac{2}{3}\Pi_0\right)\right)
\end{equation}
It is natural that $M_{\text{Jp}}>m_{\text{Jp}}$. In fact difference between these masses is because of the internal energy of the fluid. In other words we have $M_{\text{Jp}}/m_{\text{Jp}}\simeq 1+\frac{\Pi_0}{c^2}$.

Equations (\ref{mJJP}), (\ref{mJp}) and (\ref{eJp}) are the main results of this paper. So far, we have not introduced the equation of state of the fluid. In the following we study some specific equations of state and derive the stability criterion with respect to the fluid temperature and pressure of the fluid in the ideal and polytropic fluid respectively.

\subsection{Ideal fluid: $\frac{p}{\rho}=\frac{k_BT}{\mu m_H}$ }\label{ideal section}

In this section we use the ideal-fluid equation of state in order to derive a stability criterion as a function of temperature. To do so, let us introduce internal energy and the speed of sound in terms of temperature and define some dimensionless parameters. Since the fluid is assumed to be ideal, the internal energy per unit mass and the sound speed can be written as follows
\begin{equation}\label{cs ideal}
c_{\text{s}}^2=\beta \frac{k_BT}{\mu m_H},~~~\Pi_0=\frac{1}{\gamma-1}\frac{k_BT}{\mu m_H}
\end{equation}
 Where $\gamma$ is the adiabatic index, $m_H$ is the mass of hydrogen atom, $k_B$ is the Boltzmann's constant, and $T$ and $\mu$ are temperature and mean molecular weight of the fluid respectively. Furthermore $\beta=\gamma$ for the adiabatic fluids and $\beta=1$ for the isothermal case.
\begin{figure}
\centering\includegraphics[width=8.6cm]{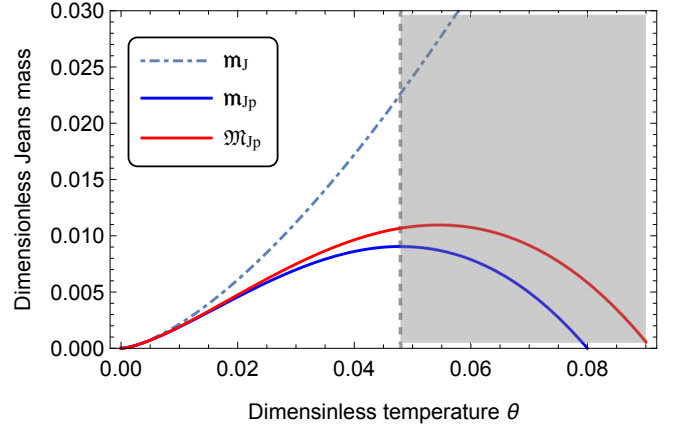}\caption{The dimensionless Jeans mass and Jeans mass-energy in terms of dimensionless temperature $\theta$. From up to down, dot-dashed curve is $\mathfrak{m}_{\text{J}}$, the red solid curve is $\mathfrak{M}_{\text{Jp}}$ and the blue solid curve corresponds to $\mathfrak{m}_{\text{Jp}}$. The vertical dashed line corresponds to $\theta_{\text{c}}=0.048$. \label{fig1}}
\end{figure}

 Therefore equation \eqref{kj} takes the following form
\begin{equation}\label{k2}
k_{\text{Jp}}^2 \simeq k_{\text{J}}^2 \left(1 + \frac{2}{c^2}\frac{k_BT}{\mu m_H}\left(\beta+\frac{\gamma}{\gamma-1}\right)\right)
\end{equation}
Consequently the difference between $k_{\text{Jp}}$ and $k_{\text{J}}$ grows with increasing temperature and reduces with increasing the mean molecular weight $\mu$. One may easily verify that in order to see a $10$\% difference between Newtonian and PN Jeans masses a temperature of order $T\sim 10^{11} K$ is required. In other words, one may look at high temperature systems in order to find a footprint of the relativistic effects on the local gravitational stability. To simplify the analysis let us also define the following dimensionless parameters
\begin{equation}\label{free dim}
W^2= \frac{\omega^2}{4\pi G\rho_0}\hspace{.25cm},\hspace{.25cm}q = \frac{kc}{\sqrt{4\pi G\rho_0}}\hspace{.25cm},\hspace{.25cm}\theta = \frac{k_BT}{\mu m_H c^2}
\end{equation}
where $W$, $q$ and $\theta$ are the dimensionless frequency, wavenumber and temperature respectively. Now the dimensionless form of the dispersion relation \eqref{w khola3} is
\begin{equation}\label{W2 vs. q theta}
W^2 = \Big(1-\frac{\gamma}{\gamma-1}\theta\Big)\,\beta\,\theta \, q^2 - \Big(1+\frac{\gamma}{\gamma-1}\theta\Big) - \frac{2}{q^2}
\end{equation}
and the dimensionless form of equation \eqref{k2} is
\begin{eqnarray}
q_{\text{Jp}}^2 \simeq q_{\text{J}}^2 + \frac{2}{\beta}\left(\beta+\frac{\gamma}{\gamma-1}\right)
\end{eqnarray}
where $q_{\text{J}}^2 = \frac{1}{\beta \,\theta}$. Now for a better comparison between Jeans masses at different temperatures let us define dimensionless masses by dividing them to the following mass parameter
\begin{equation}
\frac{\pi^{5/2}}{6}\left(\frac{c^2}{ G \rho_0^{1/3}}\right)^{3/2}
\end{equation}
The dimensionless standard Jeans mass $\mathfrak{m}_{\text{J}}$ is then
\begin{equation}\label{jeans mass dimless}
\mathfrak{m}_{\text{J}} = (\beta\,\theta)^{3/2}
\end{equation}
And the PN dimensionless Jeans mass $\mathfrak{m}_{\text{Jp}}$ is
\begin{equation}\label{PN jeans mass dimless}
\mathfrak{m}_{\text{Jp}} = \mathfrak{m}_{\text{J}} \left(1-3\left(\beta+\frac{\gamma}{\gamma-1}\right)\,\theta\right)
\end{equation}
Furthermore the PN dimensionless Jeans mass-energy $\mathfrak{M}_{\text{Jp}}$ takes the following form
\begin{equation}
\mathfrak{M}_{\text{Jp}}=  \mathfrak{m}_{\text{J}} \left(1-\left(3\beta+\frac{3\gamma-1}{\gamma-1}\right)\,\theta\right)
\end{equation}
It is clear that the fractional difference between Newtonian and PN Jeans masses grows with dimensionless temperatures as $\theta$. We have plotted these masses in Fig. \ref{fig1}. For simplicity, and without loss of generality we assume that the background fluid is monoatomic and evolves adiabatically (i.e., $\beta=\gamma$ and $\gamma=\frac{5}{3}$).
 Dashed curve is $\mathfrak{m}_{\text{J}}$, the red curve is $\mathfrak{M}_{\text{Jp}}$, and the blue curve corresponds to $\mathfrak{m}_{\text{Jp}}$. As expected at low temperatures $\theta$ or equivalently at low thermal velocities, i.e., $v_{\text{th}}\ll c$, there is no difference between Jeans masses in PN limit and Newtonian regime. However after $\theta\sim 0.01$ departure from standard case appears. The Newtonian Jeans mass increases with $\theta$ forever. However, although PN masses also increase with $\theta$, they start to decrease after a specific $\theta$. In fact $\mathfrak{m}_{\text{Jp}}$ has a maximum at $\theta=0.048$ and $\mathfrak{M}_{\text{Jp}}$ has a maximum at $\theta=0.055$. At these $\theta$ the fractional difference between PN and Newtonian Jeans masses is more than $60$\%.

However one should note that the decreasing behavior of the PN masses is not reliable. In other words at $\theta=0.048$ we have $v_{\text{th}}\sim 0.3 c$ and so the thermal velocity of the particles (or the sound speed $c_{\text{s}}$), at this temperature is of the same order of the speed of light. So the slow-motion condition is not established. One may add higher PN corrections to the hydrodynamic equations in order to study such a high temperature. Therefore let us define a critical dimensionless temperature as $\theta_{\text{c}}=0.048$ beyond which our analysis here may not be reliable. This parameter for isothermal fluid is about 0.057. So our results are limited to $\theta<\theta_{\text{c}}$. We have shown this boundary with vertical line in Fig. \ref{fig1}.

\begin{figure}
\centering\includegraphics[width=8.6cm]{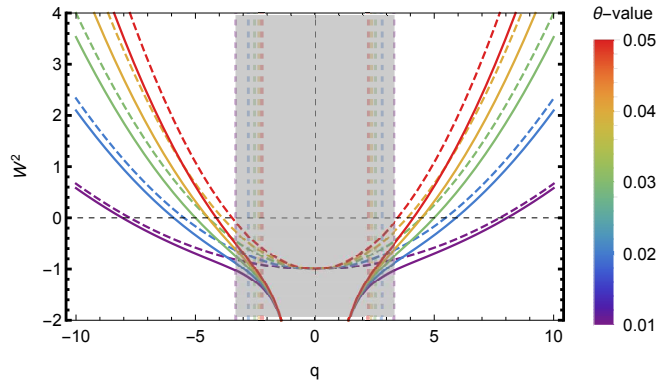}\caption{Dashed curves show the dispersion relation in Newtonian theory. From up to down, dashed curves correspond to $\theta = 0.05, 0.04, 0.03, 0.02 $ and $0.01$. The solid curves are the corresponding dispersion relation in the PN approximation. Furthermore, the vertical dashed lines show the corresponding $q_{\text{min}}$ for the above mentioned temperatures. Also, we have assumed that the adiabatic index is $\gamma=\frac{5}{3}$. \label{fig2}}
\end{figure}

Also, as we have already mentioned, in the PN approximation one can not consider the stability of the modes with very large wavelengths. In other words the wavenumber of the perturbation should satisfy the constraint \eqref{condi}. This constraint in the dimensionless form is
\begin{equation}\label{k4}
\textcolor{black}{q>\Big(\frac{2}{\beta\,\theta}\Big)^{1/4}}
 \end{equation}
Therefore according to equation (\ref{k4}), the minimum value of the dimensionless wavenumber $q$ is given by $q_{\text{min}}=(\frac{2}{\beta\,\theta})^{1/4}$. This minimum value depends on the temperature of the fluid and decreases by increasing the dimensionless temperature $\theta$.  We have shown this limit by vertical dashed lines in Fig. \ref{fig2} for temperatures in the interval $0.01<\theta<0.05$. This interval has been illustrated with a gray color area in Fig. \ref{fig2}. In this figure, we have plotted the dispersion relation in both Newtonian and PN limits at different dimensionless temperatures. It is clear that, as expected, by increasing the temperature $\theta$ the difference between dispersion relations gets larger. Also by increasing $\theta$ the instability interval, i.e., $q_{\text{min}}<q<q^*$ where $q^*$ is a wavenumber at which $W=0$ (note that in the Newtonian case $q^*=q_\text{J}$), gets smaller in both theories. This means that increasing the temperature make the instability interval shorten. One should note that this behavior is not in conflict with the fact that increasing the pressure (or equivalently the temperature) makes the Jeans mass smaller than the standard case, and consequently makes the system more unstable in comparison with the Newtonian case.

\begin{figure}
\centering\includegraphics[width=8.6cm]{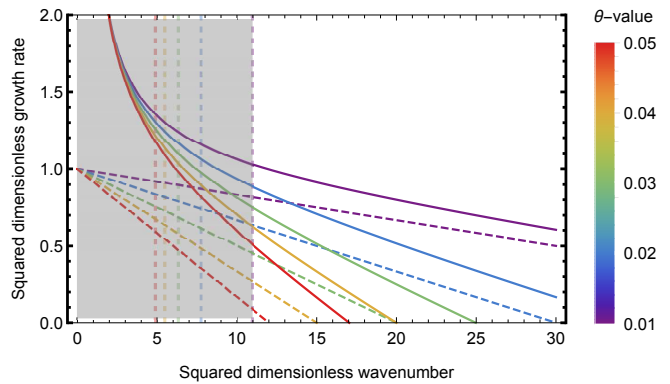}\caption{Squared growth rate $S^2$ with respect to the squared wavenumber $q^2$ for adiabatic fluid with $\gamma=\frac{5}{3}$. From up to down, dashed curves belong to the Newtonian case with $\theta =0.01, 0.02, 0.03, 0.04$ and $0.05$. Furthermore the solid curves are the corresponding curves in the PN approximation. The vertical dashed lines show $q^2_{\text{min}}$ for the above mentioned temperatures.\label{fig3}}
\end{figure}

On the other hand, at any wavenumber the oscillation frequency of the stable modes in PN approximation are less than the Newtonian case. In contrast, for unstable modes the growth rate of the perturbations in PN regime is larger than the Newtonian perturbations. In other words, PN effects not only shorten the stability interval but also enhance the instability rate. To see this behavior more clearly let us plot $S^2$ with respect to $q^2$, where $S$ is defined as $S=iW$. The result has been illustrated in Fig. \ref{fig3}.

As a final remark in this section, it should be emphasized that the Jeans stability criterion is reliable only if the Jeans wavelength is short compared to the characteristic size $L_{\text{c}}$ of the host system. Therefore this condition, e.g. $L_{\text{c}}> \lambda_{\text{Jp}}$, can be written as 
\begin{equation}\label{1n}
L_{\text{c}}>\textcolor{black}{\sqrt{\frac{\beta \pi c^2}{ G \rho_0}\theta}\left(1-\left(\beta+\frac{\gamma}{\gamma-1}\right)\theta\right)}
\end{equation}
now since the matter density is related to the number density $n$ as $\rho_0=\mu m_{H} n$,
we can rewrite the above equation as
\begin{equation}\label{cons1}
L_{\text{c}}>\textcolor{black}{5.17\times 10^{7}\sqrt{\frac{\beta\,\theta}{\mu\, n}} \left(1-\left(\beta+\frac{\gamma}{\gamma-1}\right)\theta\right)\,\text{kpc}}
\end{equation}
On the other hand, more importantly, the typical time scale for contraction of the perturbation must be smaller than the dynamical time scale $t_{\text{d}}$ of the host system.
 If without loss of generality we assume the typical free fall time is $t_{\text{ff}}\sim 1/\sqrt{G\rho_0}$ , then the following constraint should also be satisfied for triggering the perturbation growth
\begin{equation}\label{cons2}
t_{\text{d}}> 3 \text{Myr} \left(\frac{\mu \,n}{10^3\, \text{cm}^{-3}}\right)^{-1/2}
\end{equation}
It is necessary to mention that the free fall time is different from $1/\sqrt{G\rho_0}$ in the PN regime.  Therefore one may use $t_{\text{eff}}\sim 1/\sqrt{G \epsilon_0}$ as a better estimation. However, since $\epsilon_0>\rho_0$, if the constraint (\ref{cons2}) is satisfied then the corresponding constraint in the PN approximation will be satisfied.

In the Sec. \ref{Astro sec} we briefly review some high temperature systems and then check the required conditions for gravitational collapse in the PN limit and compare it with the Newtonian case. 

       \begin{table*}[!]
       \caption{Characteristics of the various astrophysical systems}\label{table1}
       \begin{center}
      \renewcommand{\arraystretch}{1.5}
         \begin{tabular}{lcccccc}
         \hline         System & $n\,(\text{cm}^{-3})$ & $T\,(\text{K})$ & Size $(\text{pc})$& $\lambda_{\text{Jp}}\,(\text{pc})  $& $100\times\Delta m_{\text{J}}/m_{\text{J}} $ \\
         \hline
         \hline
         H II regions               & $0.1-10^4$      & $10^4$ &  $0.31 - 31$          & $32.78-1.03\times10^4$      & $1.86 \times 10^{-6}$  \\
       
         NGC 7027                      & $6\times10^4$     & $3\times10^6$ & $\sim4\times 10^{-3}$ & $231.82$         & $5.59\times10^{-4}$  \\
         Cygnus loop                   & $>10^5$     & $3.5\times10^6$  & $\sim 3\times 10^{-4}$   & $193.96$         & $6.52\times10^{-4}$    \\
         ICM          & $10^{-3}$  & $10^7-10^8$ &  $\sim10^6$     & $(3.28-10.37)\times10^6$     & $(1.86-18.63)\times10^{-3}$   \\
         Fermi Bubble  & $10^{-2}$  & $10^8-10^9$   &  $10^4$   & $(3.28-10.36)\times10^6$     & $(1.86-18.63)\times10^{-2}$  \\
         HMNS         & $10^{39}$  & $10^{10}-10^{11}$ &$\sim10^{-13}$& $(1.03-3.07)\times10^{-13}$  & $(1.86-18.63)$   \\
          NDAF         & $10^{37}$  & $10^{11}$  &$\sim10^{-11}$&  $30.75\times10^{-11}$             & $18.63$                   \\
         \hline
         \end{tabular}
       \end{center} 
       \tablecomments{$n$ and $T$ are the number density and temperature respectively.  The characteristic size of the system is shown in the fourth column. For each system, we have calculated the PN Jeans wavelength in the fifth column. Furthermore the fractional difference $\Delta m_{\text{J}}/m_{\text{J}}$ has been shown for all systems in the last column. Note that for all cases we have set $\mu = 0.615$. It is worthwhile to point out that we choose thickness of the shell as a characteristic size for NGC 7027 and Cygnus loop.}
       \end{table*}

\subsection{polytropic fluid: $p=K\rho^\Gamma$}\label{polytropic fluid}
In this section we study the PN Jeans analysis for a polytropic fluid. As we know, polytropic equation of state is  widely used in modeling astrophysical systems, e.g. the interior of neutron stars, and in numerical simulations. In this case $K$ is constant of proportionality and $\Gamma$ is known as the polytropic exponent and is related to the polytropic index $n$ as $\Gamma=(n+1)/n$, here $n$ should not be confused with the number density. For a fluid system with polytropic equation of state the sound speed and the specific internal energy are given by
\begin{equation}
c_{\text{s}}^2=K \,\Gamma \rho_0^{\Gamma-1},~~~\Pi_0=\frac{K}{\Gamma-1} \rho_0^{\Gamma-1}
\end{equation}
In this case using equation \eqref{kj}, the Jeans wavelength can be written as
\begin{equation}\label{ljl}
\lambda_{\text{Jp}}=\lambda_{\text{J}}\left(1-\frac{1}{c^2}\left[\frac{\Gamma^2}{\Gamma-1}\right]K\rho_0^{\Gamma-1}\right)
\end{equation}
Furthermore the standard Jeans wavelength is written as 
\begin{equation}\label{lamdapolytropic}
\lambda_{\text{J}}=\sqrt{\frac{\pi  K \,\Gamma \rho_0^{\Gamma-2}}{G}}
\end{equation}
It is also straightforward to construct the PN Jeans mass using Eq. (\ref{ljl})
\begin{equation}\label{ljm}
m_{\text{Jp}}=m_{\text{J}}\left(1-\frac{3}{c^2}\left[\frac{\Gamma^2}{\Gamma-1}\right]K\rho_0^{\Gamma-1}\right)
\end{equation}
Interestingly for $\Gamma=2$, which is a common approximation for neutron stars, the Jeans wavelength does not directly depend on the matter density of the system and $K$ plays a central role. As in the case of ideal fluid, one can impose a constraint as $\lambda_{\text{Jp}}<L_\text{c}$ on the magnitude of the PN Jeans wavelength. Furthermore, as we have already mentioned, the typical time scale of collapse must be smaller than the dynamical time scale $t_{\text{d}}$ of the host system.

\begin{center}
\begin{figure}
\includegraphics[scale=0.86]{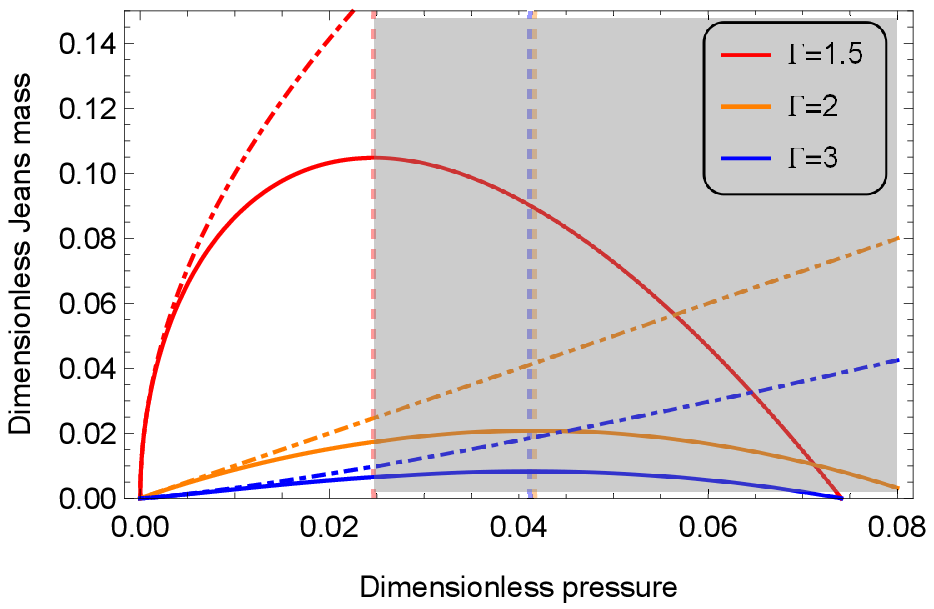}
\hspace*{.1cm}
\includegraphics[scale=0.86]{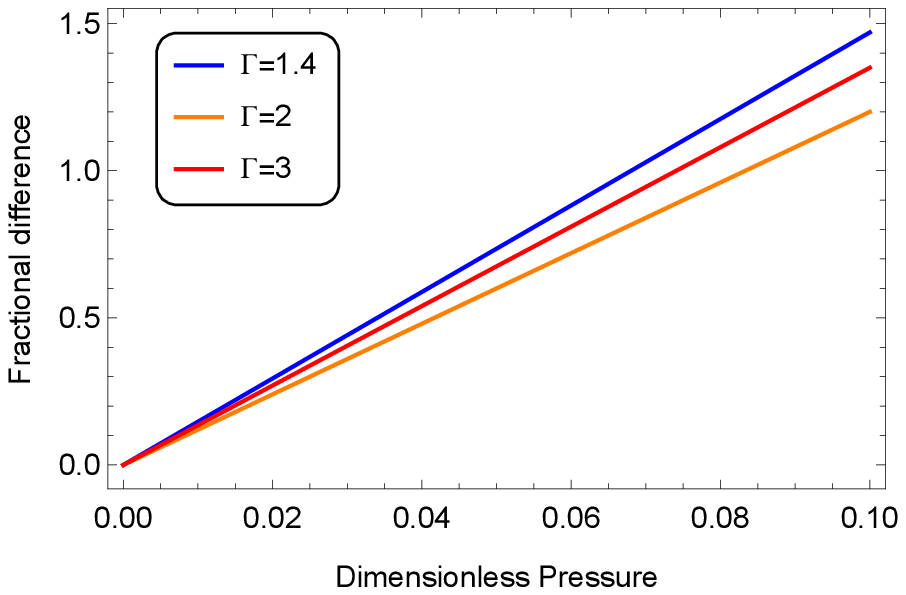}\caption{Top panel shows the Jeans masses with respect to $\mathfrak{p}$ for a polytropic fluid for different $\Gamma$. In this panel the solid curves belong to the PN case and the dot dashed curves belong to the standard case. The bottom panel shows the fractional difference between the Jeans masses.}\label{newfig}
\end{figure}
\end{center}

Similar to the previous section let us define some dimensionless parameters. In this case, for simplicity, instead of a dimensionless temperature we write the results with respect to a the dimensionless pressure $\mathfrak{p}$ 
\begin{equation}\label{dimrho}
\mathfrak{p}=\frac{K}{c^2}\,\rho_0^{\Gamma-1}
\end{equation}
Furthermore we define a dimensionless PN Jeans mass $\mathfrak{m}_{\text{Jp}}$ by dividing Eq. (\ref{ljm}) to the following mass parameter
\begin{equation}
\frac{\pi^{5/2}}{6}\left(\frac{\Gamma}{G}\right)^{3/2}\, \left(\frac{c^2}{K^{\frac{-1}{3\Gamma-4}}}\right)^{\frac{3\Gamma-4}{2\Gamma-2}}
\end{equation}
the final result is
\begin{equation}
\mathfrak{m}_{\text{\text{Jp}}}=\mathfrak{p}^{\frac{\frac{3\Gamma}{2}-2}{\Gamma-1}}\left(1-3\mathfrak{p}\frac{\Gamma^2}{\Gamma-1}\right)
\end{equation}
In fact $\mathfrak{m}_{\text{\text{Jp}}}$ has an extremum at $\mathfrak{p}_{\text{max}}=\frac{3\Gamma^2-7\Gamma+4}{3\Gamma^2\left(5\Gamma-6\right)}$. In the top panel of Fig. \ref{newfig} we have indicated $\mathfrak{p}_{\text{max}}$ with vertical dashed lines for different $\Gamma$. It is clear that beyond $\mathfrak{p}_{\text{max}}$ the PN Jeans mass starts to decrease while the Newtonian Jeans mass increases forever. As in the ideal gas case, we can not rely on our results for $\mathfrak{p}>\mathfrak{p}_{\text{max}}$. It is interesting that $\mathfrak{p}_{\text{max}}$ increases by increasing the $\Gamma$ parameter. More specifically, it is maximum at $\Gamma=2.36$ and then decreases with $\Gamma$. The fractional difference between Jeans masses with respect to $\mathfrak{p}$ has been illustrated in the bottom panel of Fig. \ref{newfig}. As expected the fractional difference $\Delta m_{\text{J}}/m_{\text{J}}$ is also sensitive to $\Gamma$. Interestingly, at a same $\mathfrak{p}$, it is minimum when $\Gamma=2$. This polytopic index is widely used for neutron stars.

As an example for this section let us calculate PN Jeans mass for a neutron star (NS). The exact form of equation of state for NSs is still not known. However the polytropic equation of state is widely used to model/simulate this type of stars. Therefore, in what follows we assume a polytropic equation of state with $\Gamma=2$ and ignore the effects of strong magnetic fields and differential rotation of material inside a NS. Thus, we find a crud estimation for the relativistic Jeans mass inside a NS, and for more reliable results one has to take into account other effects like magnetic fields and may be higher PN corrections.

It is easy to show that for a NS the PN Jeans mass start to decrease for dimensionless pressures larger than $\mathfrak{p}_{\text{max}}=0.042$. By substituting typical value of $K\simeq 0.014\, \text{m}^5\,\text{kg}^{-1}\,\text{s}^{-2}$ (\citealp{lattimer2001neutron}) into equation (\ref{dimrho}) we find the critical density $\rho_{\text{crit}}\simeq 2.7\times 10^{17} \text{kg}\,\text{m}^{-3} $.
Therefore it is easy to verify that the Newtonian gravitational potential divided by $c^2$ for a NS with  $M=1.4M_\odot$, $\rho\simeq (2.7-5.4)\times 10^{17}\, \text{kg}\,\text{m}^{-3}$, and the typical size $15\,\text{km}$ is $U/c^2\simeq M G/Rc^2\simeq 0.14$. Furthermore for this system we have $c_{\text{s}}^2/c^2\simeq 0.17$ and $\Pi_0/c^2\simeq 0.08$. As we have already mentioned these dimensionless quantities measure deviations from Newtonian gravity. It is clear that the PN corrections in this systems are significant.  

Now let us apply our PN Jeans criterion to a NS. In other words, we check the local stability of inside of a NS in the PN approximation. We use the following typical values for physical properties of NS: $\rho\simeq 5.4\times 10^{17} \,\text{kg}\text{m}^{-3}$, $p=K \rho^2$ where $K\simeq 0.014 \, \text{m}^5\,\text{kg}^{-1}\,\text{s}^{-2}$. In this case it is easy to show that the standard Jeans wavelength is $\lambda_{\text{J}}=36.32 \,\text{km}$. This wavelength is larger than the typical size of these stars, i.e., $R\simeq 15\, \text{km}$, and consequently in the Newtonian gravity the local gravitational collapse is not allowed. However in the PN limit the Jeans wavelength is $34$\% smaller. In other words we have $\lambda_{\text{\text{Jp}}}=24.11\,\text{km}$. However, we emphasize again that in order to find a more precise result one should add several physical effects which are important in the dynamics of NS.

\section{Application to high temperature environments}\label{Astro sec}
To compare the new Jeans mass with the Newtonian case for an ideal fluid, which have been calculated in Sec. \ref{ideal section}, let us first introduce some relatively high temperature environments and finally calculate the fractional difference $\Delta m_\text{J}/m_\text{J}$, where $\Delta m_{\text{J}}=m_{\text{J}}-m_{\text{Jp}}$, between Jeans masses for each system. We have summarized the results in the Tables \ref{table1} and \ref{table2}.

i) HII regions are one of the most easily observed ob-
jects in the galaxies. The sources of the ionization
in these regions are cosmic-ray protons, intense ultra violate
emission from O and early B-type stars which have
formed inside the molecular cloud, and shock waves
from supernovae regions. They are composed primarily of hydrogen. 
 The temperatures of these regions are of the order of $10^4 $ K and their size varies from a light-year to several hundred light years. For example some of them are maybe only a light-year or less across (ultra-compact HII regions) and some bigger regions (giant HII regions) are several hundred light-years across (\citealp{anderson2009molecular}). Corresponding masses of  HII regions are of the order of $10^2 -10^4 M_{\odot}$, and their typical number density in the bright parts is of the order of $0.1-10^4 \text{cm}^{-3}$. It is easy to show that Newtonian Jeans wavelength and Jeans mass for HII regions are  $1.04-32.8\times 10^4 \text{pc}$ and $2.8\times10^6-8.9\times10^8 \text{M}_\odot$ respectively.

ii) Planetary nebulae (PNe) can be considered as high temperature environments. A white dwarf that is surrounded by gas clouds can be assumed to be a planetary nebula. This kind of hot bubble can be a product of evolution of the central star from post-asymptotic giant branch star to a white dwarf. In this phase very fast and energetic winds are made by star (with speed $\sim 1000\, \text{km\,s}^{-1}$ and mass loss rates $\gtrsim 10^{-7} M_{\odot}\text{yr}^{-1}$). When such a fast wind receives to the gas cloud, there will be a shock and the gas is superheated enough to emit in X-ray (for example see \citealp{zhekov1996modelling}).

NGC 7027 is a young and dense planetary nebula. It contains a distribution of hot molecular hydrogen and ionized gas that are investigated by results of  \textit{Hubble Space Telescope} and \textit{Near-Infrared camera} and \textit{Multi Object Spectrometer} program. Density in the shell is $n_{H^+} = 6\times 10^4 \mathrm{cm^{-3}}$ and its thickness is about $0''.9 \,(1.2 \times 10^{16}\text{cm})$ (\citealp{latter2000revealing}). Plasma temperature of the ionized shell is estimated to be about $\sim 3\times10^6 \mathrm{K}$  (\citealp{kastner2001discovery}). In this case, one can show that the Newtonian Jeans wavelength is  $0.23\, \text{kpc}$ and the Newtonian Jeans mass is $5.94\times10^{9}\text{M}_\odot$.

iii) The Cygnus loop is a large soft-shelled supernova remnant in the constellation Cygnus, an emission nebula at a distance of $440\, \text{pc}$, and the angular dimensions of $2^\circ.8\times 3^\circ.5$ which corresponds to linear dimensions $21.5\times 27\, \text{pc}$ (\citealp{blair1999distance}). A thin plasma with temperature of $T\approx 3.5\times 10^6\text{K}$ is suggested as the best-fit model in \cite{kimura2013cygnus}. Furthermore this system has a thickness which is not greater than $\sim 3\times10^{-4}\,\text{pc}$ (\citealp{boss2010pulled}) and its molecular hydrogen column density is about $N_{H_2}\simeq 0.1 \times10^{21}\,\text{cm}^{-2}$ (\citealp{uyaniker2001cygnus}). Considering these values, Newtonian Jeans wavelength and Jeans mass for this system will be $0.19\, \text{kpc}$ and $5.8\times10^9\text{M}_\odot$ respectively.

iv) Rich clusters of galaxies are luminous sources of X-rays, with X-ray luminosities of $L_X = 10^{43} - 10^{45}  \mathrm{erg \; s^{-1}}$. These regions may have a dimension of several Mpc and are called as intracluster medium (ICM).  ICM has an average particle density $n \sim 10^{-3} \mathrm{cm^{-3}}$(\citealp{sarazin1986x}). Moreover the spectroscopy in X-ray indicates that this emission almost is coming from the thermal bremsstrahlung of hot plasma with a temperature of $10^7 - 10^8 \text{K}$ (\citealp{mo2010galaxy}).
  Using mentioned number density and temperature, the Newtonian Jeans wavelength and Jeans mass can be calculated as $3.3\times10^6-1.0\times10^7\, \text{pc}$ and $2.80\times10^{14}-8.86\times10^{15}\text{M}_\odot$ respectively.
  
It is worth mentioning that since the characteristic length scale of galaxy clusters is large and their density is low, one may expect that the cosmic cosmological constant $\Lambda$ plays significant role in the binding energy and consequently on the global stability of the system. The effect of $\Lambda$ on the binding energy of polytropic configurations in general relativity has been recently discussed by \cite{stuchlik2016general}. In fact, when the density of this system is low and its characteristic size is large enough, the very small repulsive cosmological constant, i.e., dark energy or vacuum energy, can sufficiently increase  the Jeans mass, see also \cite{kiessling2003jeans}. On the other hand as we mentioned, the Jeans wavelength for this system is comparable to the size of the system. Therefore, one may concern about the effects of the cosmological constant. However existence of $\Lambda$ will lead to a same contribution to both Newtonian and PN Jeans masses and consequently will decrease the fractional difference  $\Delta m_\text{J}/m_\text{J}$. However, this fraction is already small for ICM while we are looking for systems where $\Delta m_\text{J}/m_\text{J}$ is large. Therefore we do not take into account cosmological constant effects.

v) Using the Fermi Large Area Telescope two giant gamma-ray bubbles that extend nearly $10\,\text{kpc}$ in diameter have been discovered. Their positions are above and below the galactic plane (\citealp{cheng2011origin}).
 Several theoretical models have been presented to explain the formation of the Fermi Bubbles. 
 For example in \cite{mou2014fermi} it has been shown that winds from the ``post" hot accretion flow of the Sgr $A^*$ have caused inflation of the Fermi Bubble. The temperature of the gas within the Bubbles is in the range of $10^8 - 10^9\text{K}$. Moreover the number density of the Bubbles is about $\sim10^{-2} \mathrm{cm^{-3}}$ (\citealp{su2010giant}). Therefore the Newtonian Jeans wavelength is equal to $3.3\times10^6-1.0\times10^7\, \text{pc}$ and the Jeans mass is $2.8\times10^{15}-8.9\times10^{16}\text{M}_\odot$.

vi) As another high temperature system, let us mention the hyper massive neutron star (HMNS) which is a result of merger of a neuron star binary. It undergoes collapse to a rotating BH and the final result is a BH embedded in a massive, hot accretion torus containing $\sim$ 1-10\% of the total mass of the system.
Such a system is considered as a proposal for generating short-duration gamma-ray bursts. The initial temperature of the HMNS is $5 \times 10^{10} \text{K}$, and can even rise to $5 \times 10^{11}  \text{K}$ (\citealp{rezzolla2013relativistic}). The maximum density is $\rho_{\mathrm{max}}= 9 \times 10^{14} \mathrm{g\;cm^{-3}}$. Furthermore, the life time of the surrounding torus is $\sim 10\; \mathrm{ ms}$ and has a temperature of order $\gtrsim 10^{12}  \text{K}$ (\citealp{shibata2006magnetized}). Furthermore its density lies between $10^{12}$ and $10^{13} \mathrm{g\,cm^{-3}}$. The vertical and horizontal sizes of the torus are about $\sim 20 \,\mathrm{km}$ and $\sim 60\, \mathrm{km}$ respectively (\citealp{rezzolla2013relativistic}). The Newtonian Jeans wavelength and Jeans mass for the HMNS are $3.086 - 10.184\, \text{km}$ and $9\times10^{-3}-0.28\,\text{M}_\odot$ respectively.

vii) Gamma-ray Bursts (GRBs) are the most energetic phenomena in cosmological scales. In order to explain the energy source of GRBs several model have been proposed. The most popular model for generation of short GRBs is mergers of compact objects (\citealp{narayan1992gamma}). On the other hand it is commonly believed that the collapse of a single massive stars can be responsible for long GRBs (\citealp{woosley1993gamma}). For both cases, a dense and extremely high temperature disk is expected. In these hyper-accreting systems where the accretion rate is extremely high ($10^{-3} \text{M}_\odot s^{-1} \leq \dot{M} \leq 10 \text{M}_\odot s^{-1} $), the neutrino radiation is dominant cooling mechanism.  Therefore these hyper-accreting disks are usually named neutrino-dominated accretion flows (NDAFs). (e.g., \citealp{popham1999hyperaccreting,kohri2002can,Yi:2017ufn}). For this system temperature and density profile can be found in (\citealp{kohri2002can}).
One can see at  $r = 4\;r_s$ (where $r_s$ is the Schwarzschild radius) when the surface density is $10^{20} \mathrm{g\;cm^{-2}}$, and $M_{\text{BH}} = 3M_{\odot}$, the temperature and density will be $1.1 \times 10^{11}  \text{K}$ and $\rho = 2.3 \times 10^{13} \mathrm{g\;cm^{-3}}$ respectively.
Moreover this model does not work at $r\gtrsim 40\,r_s$. For simplicity we assume $L_\text{c}=40 r_\text{s}$ as a characteristic size of this system.  
 In this case the Newtonian Jeans wavelength and Jeans mass are $101.22$ km and $2.8\,\text{M}_\odot$ respectively.

\begin{table*}[t]
\caption{Properties of NDAFs at different radii}
 \begin{center}
\centering\renewcommand{\arraystretch}{1.5}
\begin{tabular}{ccccccc}
\hline
$M_{\text{BH}}$ & $r (r_s)$ & $n (10^{36} \text{cm}^{-3})$ & $\text{T}(10^{10} \text{K})$ & $m_{\text{J}} (M_{\odot})$ & $\lambda_{\text{J}} (km)$ & $\Delta m_{\text{J}} / m_{\text{J}} $\\
\hline\hline
\multirow{3}{*}{$5 M_{\odot}$}  & 4  & 24.44  & 12.73 & 1.99   & 64.11 & 0.21\\
                                  & 10 & 7.52 & 5.80 & 1.10   & 78.02 & 0.09\\
                                  & 40 & 1.27 & 1.77 & 0.45   & 105 & 0.03\\
\hline
\multirow{3}{*}{$10 M_{\odot}$}  & 4  & 32.89  & 15.52   & 2.30   & 61.02 & 0.25\\
                                  & 10 & 10.13 & 7.07 & 1.28   & 74.25 & 0.11\\
                                  & 40 & 1.70 & 2.15  & 0.52   & 99.94 & 0.04\\
\hline
\end{tabular}
 \end{center}
\tablecomments{The fractional difference between standard and the PN Jeans masses for NDAF calculated in different radii for different masses of the central BH. We have assumed that $\mu = 0.70$.\label{table2}}
\end{table*}

Now let us calculate the PN Jeans mass for the above mentioned systems. To do so, it is convenient to rewrite the Jeans masses, equations (\ref{jeans mass dimless}) and (\ref{PN jeans mass dimless}), with respect to temperature $T$ and not the dimensionless temperature $\theta$. Also it can be straightforwardly scaled to cases where temperature is $10^4  \text{K}$ and mass is $M_{\odot}$ and $n \sim 10^4\, \mathrm{cm^{-3}}$. Furthermore to have a clear form of the standard Jeans mass we use $\rho_0 = \mu m_H n$. In this case the Jeans mass in the Newtonian theory, i.e., equation (\ref{jeans mass dimless}), takes the following form
\begin{equation}\label{nje}
m_{\text{J}} \sim \frac{492}{\mu^2} \times 10^3 M_{\odot} \left(\frac{\beta\, T}{10^4  \text{K}} \right) ^{\frac{3}{2}}\left(\frac{n}{10^4\mathrm{cm^{-3}}} \right) ^{-\frac{1}{2}}
\end{equation}
Furthermore, the new Jeans mass given by equation (\ref{PN jeans mass dimless}) takes the following form
\begin{eqnarray}\label{pnje}
m_{\text{Jp}} &\sim& \frac{492}{\mu^2}\times 10^3 M_{\odot} \left( \frac{\beta\, T}{10^4\text{K}}\right) ^{\frac{3}{2}}\left(\frac{n}{10^4 \mathrm{cm^{-3}}} \right) ^{-\frac{1}{2}} \\\nonumber
&& \times\left[1-\frac{275}{\mu}\times 10^{-11}\left(\beta+\frac{\gamma }{\gamma-1}\right)\left(\frac{T}{10^4\text{K}} \right)  \right] 
\end{eqnarray} 
In the Table \ref{table1}, we have summarized some of the physical properties of the introduced high temperature astrophysical systems. Also, the Jeans mass $m_{\text{J}}$, the Jeans wave length $\lambda_{\text{J}}$, and the fractional difference $\Delta m_{\text{J}}/m_{\text{J}}$ are calculated using equations (\ref{nje}) and (\ref{pnje}). 
One can see that the fractional difference for adiabatic fluid have the following form
\begin{equation}
\frac{m_{\text{J}}-m_{\text{Jp}}}{m_{\text{J}}} = \frac{275}{\mu}\times 10^{-11} \left(\frac{\gamma^2 }{\gamma-1}\right)\left(\frac{T}{10^4 \text{K}} \right)
\end{equation}
As we mentioned before, this fraction increases linearly with temperature $T$. It is clear from Table \ref{table1} that, as expected, for systems at very high temperatures (i.e.,  $T \gtrsim 10^{11} \text{K}$) the difference between $m_{\text{J}}$ and $m_{\text{Jp}}$ is significant. More specifically this difference reaches to approximately $19$\% for HMNS and NDAF. In theses cases the PN Jeans wavelength is also comparable to the characteristic size of the system. For other cases the difference is not tangible. Therefore let us focus on these two cases. 

It is necessary to reiterate that to have a gravitational collapse, the dynamical time scale of the host system should be longer than the typical time scale for the contraction of the perturbation, see Eq. (\ref{cons2}). We know that a HMNS collapses to a BH with a torus or a supermassive neutron star (\citealp{rezzolla2013relativistic}) after $1\,\text{ms}-1\text{s}$. According to equation (\ref{cons2}) we see the dynamical time of this system is greater than its typical free fall time (where is about $\sim\,10^{-4}\text{s}$). So, the PN analysis shows that this system can be threatened by gravitational instability. It is easy to verify that for the torus the fractional difference between Jeans masses is even larger and reaches $186$\%. However this large deviation is not reliable in the sense that the temperature of the torus, i.e., $10^{12}\,\text{K}$, is larger than the critical temperature defined in Eq. (\ref{free dim}). In other words in this case one may doubt the validity of the PN analysis.

Eventually, for the last system (i.e., NDAF) the dynamical time scale is introduced as the accretion time (for more detail see \cite{kohri2002can}). In this model at $r=4r_{\text{s}}$, for BH mass $M = 3M_{\odot}$, and viscosity parameter $\alpha = 0.1$ the accretion time is estimated about $\sim 1.3\times10^{-2}\text{s}$ which is greater than $t_{\text{ff}}\sim10^{-3}\text{s}$. Thus, in principle, this location can be locally unstable. We have also obtained the temperature and number density at some different radii for two different central BHs and reported the results in the Table \ref{table2}. It is clear from this table that as expected the plasma temperature is higher at smaller radii, and consequently the fractional difference is larger. This fraction for $M = 10\,M_{\odot}$ BH is larger than a BH with mass $M = 5\,M_{\odot}$ in a same radius. This means that at least in the 1\tiny PN \normalsize approximation, such a system at the mentioned radii is probably subjected to local fragmentation. 
This fact can significantly influence the dynamics of the accretion. In fact one of the main mechanisms for transforming angular momentum throughout the accretion disk is viscosity. On the other hand local fragmentations can produce inhomogeneous pattern which in principle can exert nonzero torques to the background disk and consequently trigger the accretion flow.
It is important to mention that some features of GRB flares are consistent with a viscous accretion disk model which is locally fragmented via gravitational instability (\citealp{perna2005flares}).
From this point of view it seems that high resolution hydrodynamic simulations may confirm the existence of such local fragmentations in the accretion disks around BHs.

It is also necessary to mention that for rotating disks it is more appropriate to use the so-called Toomre criterion instead of the Jeans mass criterion. In this case it is required to find the PN version of the Toomre criterion (\citealp{kazemi2017tommre}). For rotating systems the axial and radial directions are not equivalent from the point of view of the local gravitational instability. More specifically the perturbation can propagate in both radial and axial directions. This fact makes the stability analysis complicated for the non-axisymmetric perturbations. It is well-known that there is no analytic criterion for the stability of non-axisymmetric perturbations in a differentially rotating disks. In other words Toomre criterion deals with the radially propagating axisymmetric perturbations.

\section{Conclusion}\label{Conclusion}
In this paper, we have studied the Jeans instability in 1\tiny PN\normalsize limit. In other words we have studied the first relativistic corrections to the Jeans instability. However we have idealized our analysis in the sense that we have ignored many physical effects, like magnetic fields, viscosity, radiation pressure and so on, which in principle influence the instabilities. Taking into account these effects would lead to more complete and of course to much complicated analysis.

For this purpose, we have introduced 1\tiny PN \normalsize corrections to the fluid equations. Finally linearizing the governing equations of the fluid system, we derived the gravitational instability criterion. In fact we found the new Jeans mass limit $m_{\text{Jp}}$ (\ref{mJp}). The main theoretical result of this work is that the new Jeans mass is smaller than the standard one. One may simply conclude that pressure and internal energy in the relativistic situations can in principle strengthen the gravitational force.

In the section \ref{ideal section} we applied the ideal fluid equation of state to derive a stability criterion as a function of temperature. In this system the PN Jeans mass is smaller than the Newtonian mass.

In the section \ref{polytropic fluid} we studied the Jeans instability in a polytropic fluid. Similar to the perfect fluid case the PN Jeans mass is smaller than the Newtonian mass. We have shown that there is a critical density, i.e., $\rho_{\text{crit}}$, beyond which our calculations will not be reliable. In other words higher order PN corrections are required. As an application, we found Jeans wavelength for a NS. In the PN regime the Jeans wavelength is about 34\% smaller than the Newtonian one. More specifically, in this case although the standard Jeans wavelength is larger than the radius of the star, by adding the PN corrections in this systems, the PN Jeans wavelength might be comparable to the size of the system.

In the section \ref{Astro sec}, we have calculated the new Jeans mass for some different astrophysical systems for which the ideal fluid approximation can be applied. The Newtonian and PN Jeans masses are completely similar at low temperature. However by increasing the temperature the fractional difference increases and reaches to about $19$\% in HMNS and to more than $100$\% for its torus. In the case of NDAFs, we investigated different masses for the central BH and showed that local and small disk  perturbation at small radii around BH can be seriously threatened by gravitational instabilities.

\acknowledgments
It is a pleasure to  thank Nicolas Chamel for helpful comments and discussions. M. Roshan thanks Aspen center for physics where a part of this work was carried out. Also we thank the anonymous referee for useful comments.

\bibliographystyle{apj}
\bibliography{short,PNJeans}

\begin{thebibliography}{}

\bibitem[\protect\citeauthoryear{Anderson et~al.}{Anderson
  et~al.}{2009}]{anderson2009molecular}
Anderson, L., Bania, T., Jackson, J., Clemens, D., Heyer, M., Simon, R., Shah,
  R.,  \& Rathborne, J. 2009, ApJS, 181, 255

\bibitem[\protect\citeauthoryear{Bertin \& Romeo}{Bertin \&
  Romeo}{1988}]{bertin1988global}
Bertin, G.,  \& Romeo, A.~B. 1988, A\&A, 195, 105

\bibitem[\protect\citeauthoryear{Binney \& Tremaine}{Binney \&
  Tremaine}{2008}]{binney2008galactic}
Binney, J.,  \& Tremaine, S. 2008, Galactic dynamics (Princeton university
  press)

\bibitem[\protect\citeauthoryear{Blair et~al.}{Blair
  et~al.}{1999}]{blair1999distance}
Blair, W.~P., Sankrit, R., Raymond, J.~C.,  \& Long, K.~S. 1999, AJ, 118, 942

\bibitem[\protect\citeauthoryear{Blanchet}{Blanchet}{2006}]{blanchet2006gravitational}
Blanchet, L. 2006, LRR, 9, 4

\bibitem[\protect\citeauthoryear{Blandford \& Pringle}{Blandford \&
  Pringle}{1976}]{blandford1976kelvin}
Blandford, R.,  \& Pringle, J. 1976, MNRAS, 176, 443

\bibitem[\protect\citeauthoryear{Blandford \& Teukolsky}{Blandford \&
  Teukolsky}{1976}]{blandford1976arrival}
Blandford, R.,  \& Teukolsky, S.~A. 1976, ApJ, 205, 580

\bibitem[\protect\citeauthoryear{Boss \& Keiser}{Boss \&
  Keiser}{2010}]{boss2010pulled}
Boss, A.~P.,  \& Keiser, S.~A. 2010, ApJ, 717, L1

\bibitem[\protect\citeauthoryear{Burke}{Burke}{1971}]{burke1971gravitational}
Burke, W.~L. 1971, JMP, 12, 401

\bibitem[\protect\citeauthoryear{Capozziello et~al.}{Capozziello
  et~al.}{2012}]{capozziello2012jeans}
Capozziello, S., De~Laurentis, M., De~Martino, I., Formisano, M.,  \& Odintsov,
  S. 2012, PhRvD, 85, 044022

\bibitem[\protect\citeauthoryear{Chandrasekhar}{Chandrasekhar}{1965}]{chandrasekhar1965post}
Chandrasekhar, S. 1965, ApJ, 142, 1488

\bibitem[\protect\citeauthoryear{Chandrasekhar}{Chandrasekhar}{1967}]{chandrasekhar1967post}
Chandrasekhar, S. 1967, ApJ, 148, 621

\bibitem[\protect\citeauthoryear{Chandrasekhar}{Chandrasekhar}{1969}]{chandrasekhar1969conservation}
Chandrasekhar, S. 1969, ApJ, 158, 45

\bibitem[\protect\citeauthoryear{Chandrasekhar \& Esposito}{Chandrasekhar \&
  Esposito}{1970}]{chandrasekhar1970212}
Chandrasekhar, S.,  \& Esposito, F.~P. 1970, ApJ, 160, 153

\bibitem[\protect\citeauthoryear{Chandrasekhar \& Nutku}{Chandrasekhar \&
  Nutku}{1969}]{chandrasekhar1969Second}
Chandrasekhar, S.,  \& Nutku, Y. 1969, ApJ, 158, 55

\bibitem[\protect\citeauthoryear{Cheng et~al.}{Cheng
  et~al.}{2011}]{cheng2011origin}
Cheng, K.-S., Chernyshov, D., Dogiel, V., Ko, C.-M.,  \& Ip, W.-H. 2011, ApJ,
  731, L17

\bibitem[\protect\citeauthoryear{Damour \& Taylor}{Damour \&
  Taylor}{1991}]{damour1991orbital}
Damour, T.,  \& Taylor, J.~H. 1991, ApJ, 366, 501

\bibitem[\protect\citeauthoryear{Demianski \& Ivanov}{Demianski \&
  Ivanov}{1997}]{demianski1997dynamics}
Demianski, M.,  \& Ivanov, P. 1997, A\&A, 324, 829

\bibitem[\protect\citeauthoryear{Elmegreen}{Elmegreen}{1995}]{elmegreen1995effective}
Elmegreen, B. 1995, MNRAS, 275, 944

\bibitem[\protect\citeauthoryear{Elmegreen}{Elmegreen}{1987}]{elmegreen1987supercloud}
Elmegreen, B.~G. 1987, ApJ, 312, 626

\bibitem[\protect\citeauthoryear{Elmegreen}{Elmegreen}{2011}]{elmegreen2011gravitational}
Elmegreen, B.~G. 2011, ApJ, 737, 10

\bibitem[\protect\citeauthoryear{Epstein}{Epstein}{1977}]{epstein1977binary}
Epstein, R. 1977, ApJ, 216, 92

\bibitem[\protect\citeauthoryear{Faber \& Rasio}{Faber \&
  Rasio}{2000}]{faber2000post}
Faber, J.~A.,  \& Rasio, F.~A. 2000, PhRvD, 62, 064012

\bibitem[\protect\citeauthoryear{Ferrari, Trussoni, \& Zaninetti}{Ferrari
  et~al.}{1978}]{ferrari1978relativistic}
Ferrari, A., Trussoni, E.,  \& Zaninetti, L. 1978, A\&A, 64, 43

\bibitem[\protect\citeauthoryear{Freundlich, Jog, \& Combes}{Freundlich
  et~al.}{2014}]{freundlich2014local}
Freundlich, J., Jog, C.~J.,  \& Combes, F. 2014, A\&A, 564, A7

\bibitem[\protect\citeauthoryear{Gammie}{Gammie}{1996}]{gammie1996linear}
Gammie, C.~F. 1996, ApJ, 462, 725

\bibitem[\protect\citeauthoryear{Hardee \& Norman}{Hardee \&
  Norman}{1988}]{hardee1988spatial}
Hardee, P.~E.,  \& Norman, M.~L. 1988, ApJ, 334, 70

\bibitem[\protect\citeauthoryear{Hobson, Efstathiou, \& Lasenby}{Hobson
  et~al.}{2006}]{hobson2006general}
Hobson, M.~P., Efstathiou, G.~P.,  \& Lasenby, A.~N. 2006, General relativity:
  an introduction for physicists (Cambridge University Press)

\bibitem[\protect\citeauthoryear{Hosseinirad et~al.}{Hosseinirad
  et~al.}{2017}]{hosseinirad2017gravitational}
Hosseinirad, M., Naficy, K., Abbassi, S.,  \& Roshan, M. 2017, MNRAS, 465, 1645; arXiv:1611.00139 [astro-ph.GA]

\bibitem[\protect\citeauthoryear{Hulse \& Taylor}{Hulse \&
  Taylor}{1975}]{hulse1975Discovery}
Hulse, R.~A.,  \& Taylor, J.~H. 1975, ApJ, 195, L51

\bibitem[\protect\citeauthoryear{Jog}{Jog}{1996}]{jog1996local}
Jog, C.~J. 1996, MNRAS, 278, 209

\bibitem[\protect\citeauthoryear{Jog}{Jog}{2013}]{jog2013jeans}
Jog, C.~J. 2013, MNRAS, 434, L56

\bibitem[\protect\citeauthoryear{Jog \& Solomon}{Jog \&
  Solomon}{1984}]{jog1984two}
Jog, C.~J.,  \& Solomon, P. 1984, ApJ, 276, 114

\bibitem[\protect\citeauthoryear{Kastner, Vrtilek, \& Soker}{Kastner
  et~al.}{2001}]{kastner2001discovery}
Kastner, J.~H., Vrtilek, S.~D.,  \& Soker, N. 2001, ApJ, 550, L189

\bibitem[\protect\citeauthoryear{Kato}{Kato}{1972}]{kato1972oscillation}
Kato, S. 1972, PASJ, 24, 61

\bibitem[\protect\citeauthoryear{Kazemi, Nazari, \& Roshan}{Kazemi
  et~al.}{2017}]{kazemi2017tommre}
Kazemi, A., Nazari, E.,  \& Roshan, M. 2017, Toomre criterion in post-newtonian
  theory, work in progress

\bibitem[\protect\citeauthoryear{Kiessling}{Kiessling}{2003}]{kiessling2003jeans}
Kiessling, M. K.-H. 2003, Adv. Appl. Math., 31, 132

\bibitem[\protect\citeauthoryear{Kim \& Ostriker}{Kim \&
  Ostriker}{2001}]{kim2001amplification}
Kim, W.-T.,  \& Ostriker, E.~C. 2001, ApJ, 559, 70

\bibitem[\protect\citeauthoryear{Kimura et~al.}{Kimura
  et~al.}{2013}]{kimura2013cygnus}
Kimura, M., Tsunemi, H., Tomida, H., Sugizaki, M., Ueno, S., Hanayama, T.,
  Yoshidome, K.,  \& Sasaki, M. 2013, PASJ, 65, 14

\bibitem[\protect\citeauthoryear{Kohri \& Mineshige}{Kohri \&
  Mineshige}{2002}]{kohri2002can}
Kohri, K.,  \& Mineshige, S. 2002, ApJ, 577, 311

\bibitem[\protect\citeauthoryear{Lai \& Wiseman}{Lai \&
  Wiseman}{1996}]{lai1996innermost}
Lai, D.,  \& Wiseman, A.~G. 1996, PhRvD, 54, 3958

\bibitem[\protect\citeauthoryear{Latter et~al.}{Latter
  et~al.}{2000}]{latter2000revealing}
Latter, W.~B., Dayal, A., Bieging, J.~H., Meakin, C., Hora, J.~L., Kelly,
  D.~M.,  \& Tielens, A. 2000, ApJ, 539, 783

\bibitem[\protect\citeauthoryear{Lattimer \& Prakash}{Lattimer \&
  Prakash}{2001}]{lattimer2001neutron}
Lattimer, J.,  \& Prakash, M. 2001, ApJ, 550, 426

\bibitem[\protect\citeauthoryear{Mo, Van~den Bosch, \& White}{Mo
  et~al.}{2010}]{mo2010galaxy}
Mo, H., Van~den Bosch, F.,  \& White, S. 2010, Galaxy formation and evolution
  (Cambridge University Press)

\bibitem[\protect\citeauthoryear{Mou et~al.}{Mou et~al.}{2014}]{mou2014fermi}
Mou, G., Yuan, F., Bu, D., Sun, M.,  \& Su, M. 2014, ApJ, 790, 109

\bibitem[\protect\citeauthoryear{Narayan, Paczynski, \& Piran}{Narayan
  et~al.}{1992}]{narayan1992gamma}
Narayan, R., Paczynski, B.,  \& Piran, T. 1992, ApJ, 395, L83

\bibitem[\protect\citeauthoryear{Pati \& Will}{Pati \&
  Will}{2000}]{pati2000post}
Pati, M.~E.,  \& Will, C.~M. 2000, PhRvD, 62, 124015

\bibitem[\protect\citeauthoryear{Perna, Armitage, \& Zhang}{Perna
  et~al.}{2005}]{perna2005flares}
Perna, R., Armitage, P.~J.,  \& Zhang, B. 2005, ApJ, 636, L29

\bibitem[\protect\citeauthoryear{Poisson \& Will}{Poisson \&
  Will}{2014}]{poisson2014gravity}
Poisson, E.,  \& Will, C.~M. 2014, Gravity: Newtonian, Post-Newtonian,
  Relativistic (Cambridge University Press)

\bibitem[\protect\citeauthoryear{Popham, Woosley, \& Fryer}{Popham
  et~al.}{1999}]{popham1999hyperaccreting}
Popham, R., Woosley, S.~E.,  \& Fryer, C. 1999, ApJ, 518, 356

\bibitem[\protect\citeauthoryear{Rafikov}{Rafikov}{2001}]{rafikov2001local}
Rafikov, R.~R. 2001, MNRAS, 323, 445

\bibitem[\protect\citeauthoryear{Rezzolla \& Zanotti}{Rezzolla \&
  Zanotti}{2013}]{rezzolla2013relativistic}
Rezzolla, L.,  \& Zanotti, O. 2013, Relativistic hydrodynamics (Oxford
  University Press)

\bibitem[\protect\citeauthoryear{Romeo}{Romeo}{1992}]{romeo1992stability}
Romeo, A.~B. 1992, MNRAS, 256, 307

\bibitem[\protect\citeauthoryear{Romeo \& Wiegert}{Romeo \&
  Wiegert}{2011}]{romeo2011effective}
Romeo, A.~B.,  \& Wiegert, J. 2011, MNRAS, 416, 1191

\bibitem[\protect\citeauthoryear{Roshan}{Roshan}{2012}]{roshan2012parametrized}
Roshan, M. 2012, CQGra, 29, 215001; arXiv:1208.3184 [gr-qc]

\bibitem[\protect\citeauthoryear{Roshan \& Abbassi}{Roshan \&
  Abbassi}{2014}]{roshan2014jeans}
Roshan, M.,  \& Abbassi, S. 2014, PhRvD, 90, 044010; arXiv:1407.6431 [astro-ph.GA]

\bibitem[\protect\citeauthoryear{Roshan \& Abbassi}{Roshan \&
  Abbassi}{2015a}]{roshan2015local}
Roshan, M.,  \& Abbassi, S. 2015a, Ap\&SS, 358, 1; arXiv:1506.00942 [astro-ph.GA]

\bibitem[\protect\citeauthoryear{Roshan \& Abbassi}{Roshan \&
  Abbassi}{2015b}]{roshan2015stability}
Roshan, M.,  \& Abbassi, S. 2015b, ApJ, 802, 9; arXiv:1501.04715 [astro-ph.GA]

\bibitem[\protect\citeauthoryear{Sarazin}{Sarazin}{1986}]{sarazin1986x}
Sarazin, C.~L. 1986, RvMP, 58, 1

\bibitem[\protect\citeauthoryear{Shadmehri \& Khajenabi}{Shadmehri \&
  Khajenabi}{2012}]{shadmehri2012gravitational}
Shadmehri, M.,  \& Khajenabi, F. 2012, MNRAS, 421, 841

\bibitem[\protect\citeauthoryear{Shibata et~al.}{Shibata
  et~al.}{2006}]{shibata2006magnetized}
Shibata, M., Duez, M.~D., Liu, Y.~T., Shapiro, S.~L.,  \& Stephens, B.~C. 2006,
  PhRvL, 96, 031102

\bibitem[\protect\citeauthoryear{Siegel et~al.}{Siegel
  et~al.}{2013}]{siegel2013magnetorotational}
Siegel, D.~M., Ciolfi, R., Harte, A.~I.,  \& Rezzolla, L. 2013, PhRvD, 87,
  121302

\bibitem[\protect\citeauthoryear{Stuchl{\'\i}k, Hled{\'\i}k, \&
  Novotn{\`y}}{Stuchl{\'\i}k et~al.}{2016}]{stuchlik2016general}
Stuchl{\'\i}k, Z., Hled{\'\i}k, S.,  \& Novotn{\`y}, J. 2016, PhRvD, 94, 103513

\bibitem[\protect\citeauthoryear{Su, Slatyer, \& Finkbeiner}{Su
  et~al.}{2010}]{su2010giant}
Su, M., Slatyer, T.~R.,  \& Finkbeiner, D.~P. 2010, ApJ, 724, 1044

\bibitem[\protect\citeauthoryear{Toomre}{Toomre}{1964}]{toomre1964gravitational}
Toomre, A. 1964, ApJ, 139, 1217

\bibitem[\protect\citeauthoryear{Uyan{\i}ker et~al.}{Uyan{\i}ker
  et~al.}{2001}]{uyaniker2001cygnus}
Uyan{\i}ker, B., F{\"u}rst, E., Reich, W., Aschenbach, B.,  \& Wielebinski, R.
  2001, A\&A, 371, 675

\bibitem[\protect\citeauthoryear{Vandervoort}{Vandervoort}{1970}]{vandervoort1970density}
Vandervoort, P.~O. 1970, ApJ, 161, 87

\bibitem[\protect\citeauthoryear{Wang \& Silk}{Wang \&
  Silk}{1994}]{wang1994gravitational}
Wang, B.,  \& Silk, J. 1994, ApJ, 427, 759

\bibitem[\protect\citeauthoryear{Wielgus et~al.}{Wielgus
  et~al.}{2015}]{wielgus2015local}
Wielgus, M., Fragile, P.~C., Wang, Z.,  \& Wilson, J. 2015, MNRAS, 447, 3593

\bibitem[\protect\citeauthoryear{Will}{Will}{1994}]{will1994proceedings}
Will, C. 1994, in Proceedings of the Eighth Nishinomiya-Yukawa Memorial
  Symposium on Relativistic Cosmology, ed. M.~Sasaki (Tokyo: Universal Academy
  Press)

\bibitem[\protect\citeauthoryear{Will}{Will}{1987}]{thorne1987300}
Will, C.~M. 1987, in 300 years of gravitation, ed. K.~Thorne, S.~Hawking, \&
  W.~Israel (Cambridge, England: Cambridge University Press), 80

\bibitem[\protect\citeauthoryear{Will}{Will}{2011}]{will2011unreasonable}
Will, C.~M. 2011, PNAS, 108, 5938

\bibitem[\protect\citeauthoryear{Will}{Will}{2014}]{will2014confrontation}
Will, C.~M. 2014, LRR, 17, 4

\bibitem[\protect\citeauthoryear{Woosley}{Woosley}{1993}]{woosley1993gamma}
Woosley, S. 1993, ApJ, 405, 273

\bibitem[\protect\citeauthoryear{Yi et~al.}{Yi et~al.}{2017}]{Yi:2017ufn}
Yi, T., Gu, W.-M., Yuan, F., Liu, T.,  \& Mu, H.-J. 2017, arXiv:1701.07573

\bibitem[\protect\citeauthoryear{Zhang, MacFadyen, \& Wang}{Zhang
  et~al.}{2009}]{zhang2009three}
Zhang, W., MacFadyen, A.,  \& Wang, P. 2009, ApJ, 692, L40

\bibitem[\protect\citeauthoryear{Zhekov \& Perinotto}{Zhekov \&
  Perinotto}{1996}]{zhekov1996modelling}
Zhekov, S.,  \& Perinotto, M. 1996, A\&A, 309, 648

\end{thebibliography}
\clearpage
\end{document}